\pdfoutput=1
\documentclass[reprint,onecolumn,amsmath,amssymb,jap,aip,groupedaddress,nofootinbib]{revtex4-1}
\usepackage{graphicx}
\usepackage{xcolor}
\usepackage{url}
\usepackage{siunitx}
\usepackage{xspace}
\usepackage{hyperref}

\DeclareSIUnit{\micm}{\micro\metre}

\usepackage[english]{babel}
\newcommand{\etal}{\textit{et al.}\xspace}

\usepackage{subfigure}

\def\figurewidth{0.7\textwidth}

\def\@pdfauthor{W.W.Koelmans,J.B.C.Engelen,L.Abelmann}
\def\@pdftitle{Probe-base data storage}
\def\@pdfsubject{Probe storage review}
\def\@pdfkeywords{probe storage, atomic storage, nanopositioning, cantilever array, topographic, thermomechanical, phase-change material, magnetic, ferroelectric, molecular storage, data density}
\begin{document}

\title{Probe-based data storage}

\author{Wabe W. Koelmans}
\altaffiliation{Current affiliation: IBM Research -- Zurich, R\"uschlikon, Switzerland}
\email{wko@zurich.ibm.com}
\affiliation{MESA$^+$ Institute for Nanotechnology, University of Twente, Enschede, the Netherlands.}

\author{Johan B. C.~Engelen}
\altaffiliation{Current affiliation: IBM Research -- Zurich, R\"uschlikon, Switzerland}
\affiliation{MESA$^+$ Institute for Nanotechnology, University of Twente, Enschede, the Netherlands.}

\author{Leon Abelmann}
\altaffiliation{Also at KIST Europe, Saarbr\"ucken, Germany}
\affiliation{MESA$^+$ Institute for Nanotechnology, University of Twente, Enschede, the Netherlands.}

% \author[mesa,ibm]{W.W.~Koelmans}
% \ead{wko@zurich.ibm.com}
%
% \author[mesa,ibm]{J.B.C.~Engelen}
% \ead{jen@zurich.ibm.com}
%
% \author[mesa,kist]{L.~Abelmann\corref{cor1}}
% \ead{l.abelmann@utwente.nl}
%
% \cortext[cor1]{Corresponding author}
% 
% \address[mesa]{MESA+ Institute for Nanotechnology, University of Twente, Enschede, the Netherlands}
% \address[ibm]{IBM Research -- Zurich, R\"uschlikon, Switzerland}
% \address[kist]{KIST Europe, Saarbr\"ucken, Germany}

\begin{abstract}
  Probe-based data storage attracted many researchers from academia
  and industry, resulting in unprecedented high data-density
  demonstrations. This topical review gives a comprehensive overview
  of the main contributions that led to the major accomplishments in
  probe-based data storage. The most investigated technologies are
  reviewed: topographic, phase-change, magnetic, ferroelectric and
  atomic and molecular storage. Also, the positioning of probes and
  recording media, the cantilever arrays and parallel readout of the
  arrays of cantilevers are discussed.  This overview serves two
  purposes. First, it provides an overview for new
  researchers entering the field of probe storage, as probe storage seems to be the
  only way to achieve data storage at atomic densities. Secondly, 
  there is an enormous wealth of invaluable findings 
  that can also be applied to many other fields of nanoscale research such as
  probe-based nanolithography, 3D~nanopatterning, solid-state memory
  technologies and ultrafast probe microscopy.
\end{abstract}

\maketitle

%\begin{keyword}
%%% keywords here, in the form: keyword \sep keyword
%probe storage \sep atomic storage \sep nanopositioning \sep cantilever array \sep topographic \sep thermomechanical \sep phase change material \sep magnetic \sep ferroelectric \sep molecular \sep data density
%%% PACS codes here, in the form: \PACS code \sep code
%
%%% MSC codes here, in the form: \MSC code \sep code
%%% or \MSC[2008] code \sep code (2000 is the default)
%
%\end{keyword}
%
%\end{frontmatter}

%#################################################################
\section{Introduction}
For more than 50~years, the search for technologies that offer ever
increasing densities of digital data storage has been successful. Of
all solutions, probe-based data storage has attracted a lot of
attention over the past decades. In this overview, a wide variety of
probe storage implementations is discussed and many initial
experiments are shown. A few implementations have been matured further,
and, in the case of thermomechanical storage, this led to a first prototype in 2005\footnote{http://www.physorg.com/news3361.html, last accessed at 21 March, 2015.} 
with, for that time, revolutionary areal densities of~\SI{1}{Tb~in^{-2}}. Since then, demonstrations of much higher densities have been published, showing that probe storage can outperform any other storage technology.

Probe storage is attractive because the bit size is not determined by the maximum resolution of lithographical processes, which become increasingly costly: Probes can be chemically etched and have the potential to be atomically sharp without expensive manufacturing steps.

The main challenge of probe storage is to scale up a single probe operated under laboratory conditions to large probe arrays working at high speeds in consumer products. When increasing the number of probes, the positioning accuracy has to be maintained, also under externally applied shocks and ambient-temperature variations. Fabrication-induced deviations between the probes in the array have to be minimized to ensure that all probes function correctly and remain working throughout the device life-time. The medium and the tips have to endure many read-write cycles, and the tips must be able to travel for kilometers over the storage medium.

\subsection{Schematic of probe-storage system}
\begin{figure}
  \centering
  \includegraphics[width=10cm]{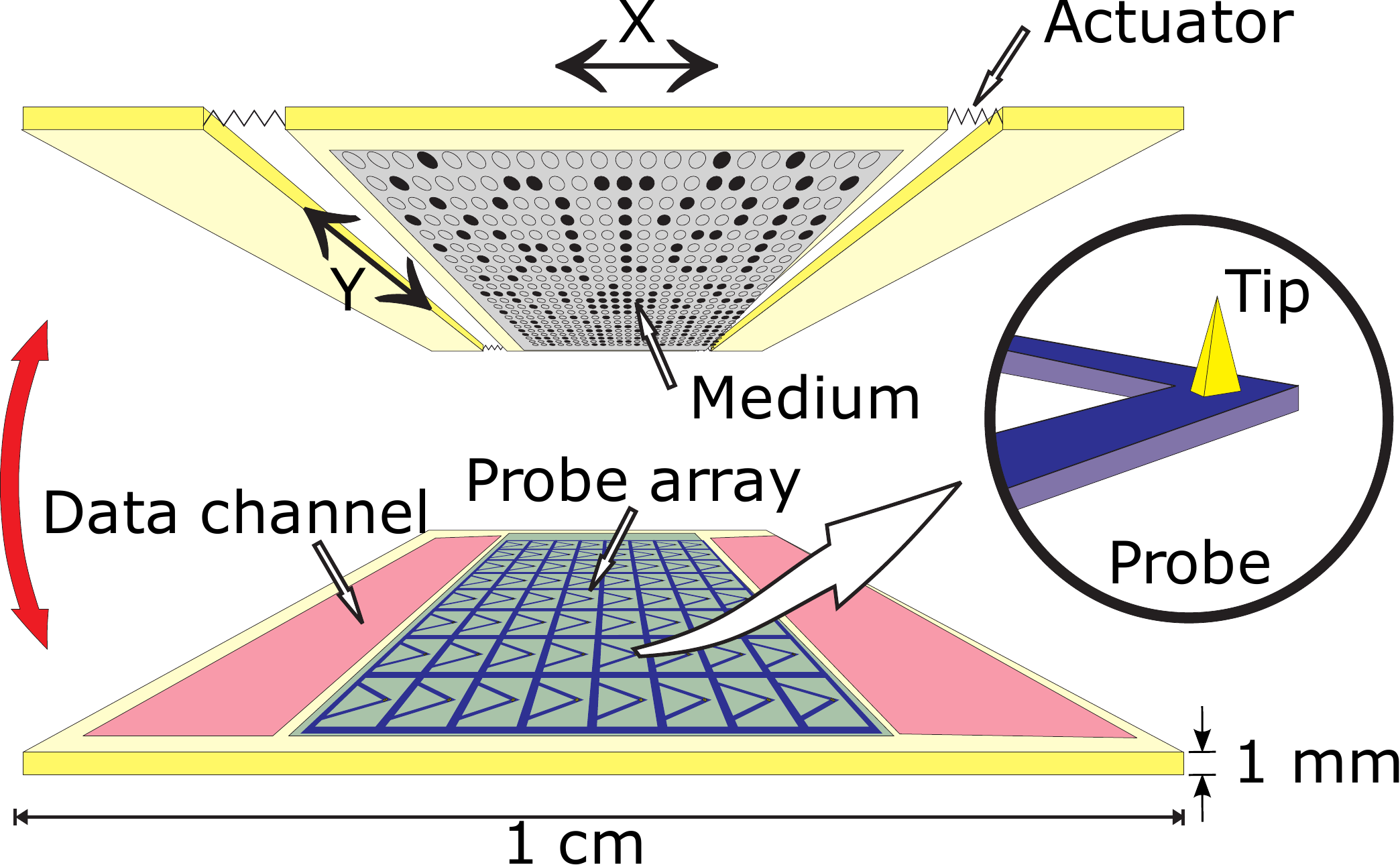}
  \caption{Schematic of a probe-based recording system.}
  \label{fig:architecture}
\end{figure}

A schematic of an architecture for a storage device based on probe technology is shown in Figure~\ref{fig:architecture}. This kind of architecture was first proposed by IBM~\cite{Lutwyche1999}. The core of the storage device is an array of probes with a moving medium on the opposite side. Each probe can locally alter a property of the medium to write a bit. Reading is accomplished by the same probe that wrote the bit.

A variation on the architecture shown in Figure~\ref{fig:architecture} uses a spinning disk, as in hard-disk drives, to position the medium. Such a design offers constant high positioning speeds. Spinning disks are mainly researched in combination with ferroelectric probe storage~\cite{Zhao2008, Hiranaga2009}.

\subsection{Current status of probe storage}
The maturity of the various types of probe storage differs significantly. Four popular types, namely phase change, magnetic, thermomechanical and ferroelectric storage are listed in table~\ref{tab:storage}.

%\ctable[
%  caption = Achievements of the most mature probe-storage technologies.,
%  label = tab:storage,
%  pos = htbp,
%]{@{}lllll@{}}{
%}{\toprule
%        & Phase-change &  Magnetic & Thermomech. & Ferroelectric \\
%   \cmidrule(lr){2-2}\cmidrule(lr){3-3}\cmidrule(l){4-4}\cmidrule(lr){5-5}
% Density  & \SI{3.3}{Tb/in^2}~\cite{Hamann2006} & \SI{60}{Gb/in^2}~\cite{Kappenberger2009} & \SI{4.0}{Tb/in^2}~\cite{Wiesmann2009} & \SI{4.0}{Tb/in^2}~\cite{Tanaka2010}\\
% Est. Max. density & $\approx$\SI{10}{Tb/in^2}~\cite{Wright2006} & $\approx$\SI{100}{Tb/in^2}~\cite{Kryder2008}& $\approx$\SI{10}{Tb/in^2}~\cite{Wiesmann2009} & $>$\SI{10}{Tb/in^2}~\cite{Cho2005}\\
% Read speed per probe & \SI{50}{Mb.s^{-1}}~\cite{Hamann2006} &  $<$ \SI{10}{b/s}~\cite{Onoue2008}& \SI{40}{kb.s^{-1}}~\cite{Sebastian2009} & \SI{2}{Mb.s^{-1}}~\cite{Hiranaga2009}\\
% Write speed per probe & \SI{50}{Mb.s^{-1}}~\cite{Hamann2006} &  $<$ \SI{10}{b/s}~\cite{Onoue2008}  & \SI{1}{Mb.s^{-1}}~\cite{Cannara2008} & \SI{50}{kb.s^{-1}}\cite{Cho2006}\\
%Travel per probe & \SI{2.5}{m}~\cite{Bhaskaran2009} &  \SI{0.5}{m}~\cite{Onoue2008}  & \SI{750}{m}~\cite{Lantz2009} & \SI{5000}{m}~\cite{Tayebi2010}\\
%\bottomrule
%}

\begin{table}
\caption{\label{tab:storage}Achievements of the most mature probe-storage technologies.}
\begin{ruledtabular}
\begin{tabular}{@{}lllll@{}}
        & Phase-change &  Magnetic & Thermomech. & Ferroelectric \\
\hline
%   \crule(lr){2-2}\crule(lr){3-3}\crule(l){4-4}\crule(lr){5-5}
 Density  & \SI{3.3}{Tb/in^2}~\cite{Hamann2006} & \SI{60}{Gb/in^2}~\cite{Kappenberger2009} & \SI{4.0}{Tb/in^2}~\cite{Wiesmann2009} & \SI{4.0}{Tb/in^2}~\cite{Tanaka2010}\\
 Est. Max. density & $\approx$\SI{10}{Tb/in^2}~\cite{Wright2006} & $\approx$\SI{100}{Tb/in^2}~\cite{Kryder2008}& $\approx$\SI{10}{Tb/in^2}~\cite{Wiesmann2009} & $>$\SI{10}{Tb/in^2}~\cite{Cho2005}\\
 Read speed per probe & \SI{50}{Mb.s^{-1}}~\cite{Hamann2006} &  $<$ \SI{10}{b/s}~\cite{Onoue2008}& \SI{40}{kb.s^{-1}}~\cite{Sebastian2009} & \SI{2}{Mb.s^{-1}}~\cite{Hiranaga2009}\\
 Write speed per probe & \SI{50}{Mb.s^{-1}}~\cite{Hamann2006} &  $<$ \SI{10}{b/s}~\cite{Onoue2008}  & \SI{1}{Mb.s^{-1}}~\cite{Cannara2008} & \SI{50}{kb.s^{-1}}\cite{Cho2006}\\
Travel per probe & \SI{2.5}{m}~\cite{Bhaskaran2009} &  \SI{0.5}{m}~\cite{Onoue2008}  & \SI{750}{m}~\cite{Lantz2009} & \SI{5000}{m}~\cite{Tayebi2010}\\
\end{tabular}
\end{ruledtabular}
\end{table}

Probe-based data storage has attracted much scientific and commercial
interest~\cite{Yang2007b, Wright2011, ProbeStorageChap2011}. After more than two
decades of research on probe storage, an overview of what has been
accomplished so far, together with an outlook of the potential of
probe storage, is presented in this overview. The time seems right to
assess the results because most large industrial efforts devoted to
probe storage have been discontinued in the past years. The focus has
shifted towards new applications that exploit the technology
developed for probe storage. Examples of such applications are probe
lithography~\cite{Garcia2014}, nanofabrication and
3D~nanopatterning~\cite{Pires2010}. 

\begin{figure}
  \centering
  \includegraphics[width=\figurewidth]{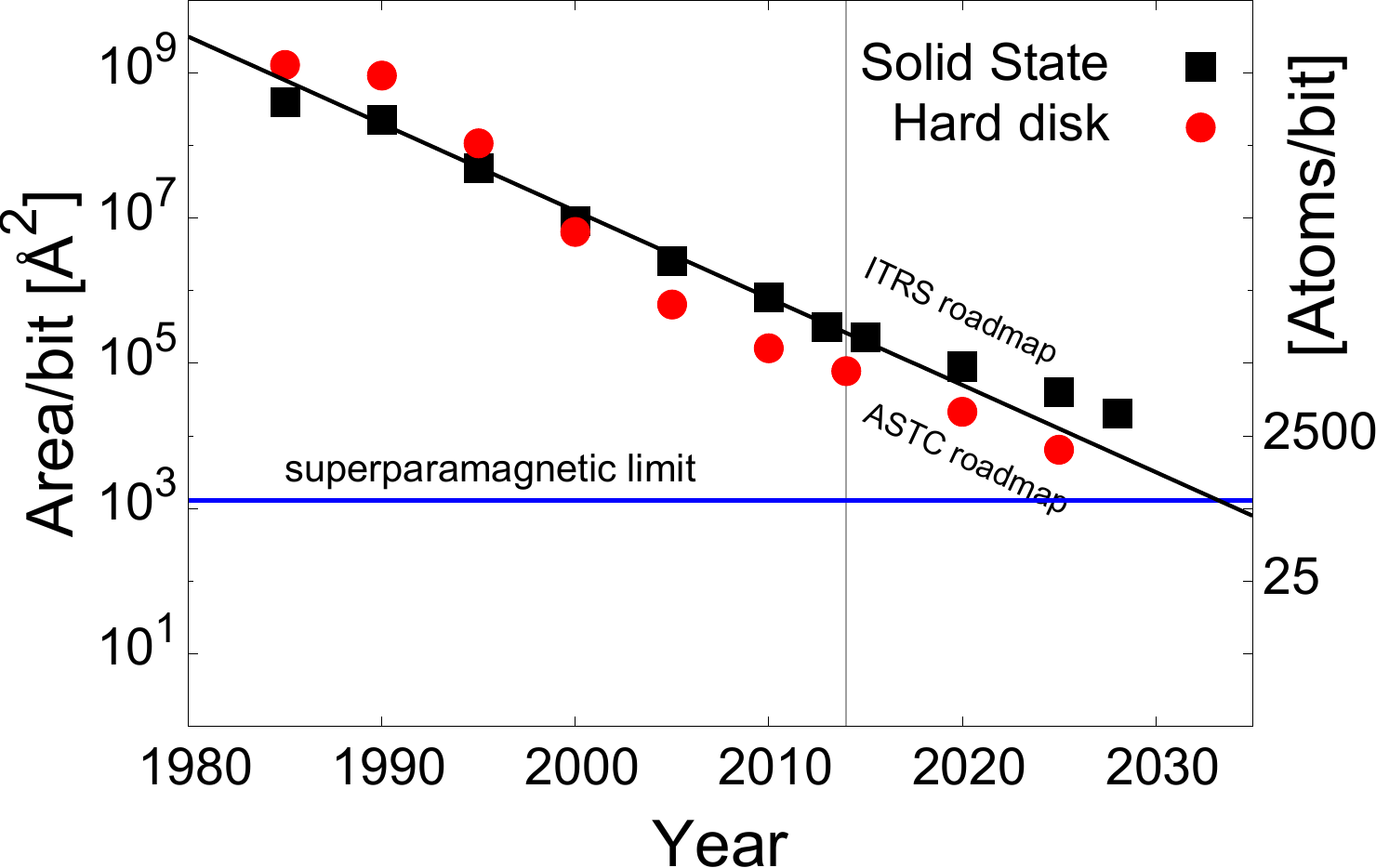}
  \caption{Reduction in the size of transistors (solid-state memories)
    or bits (hard disk). The first fundamental limit that will be
    reached is thermal stability (super-paramagnetism) in magnetic
    data storage. Probe storage provides a viable rout to higher
    densities, all the way to single-atom storage.}
  \label{fig:AngstromsPerBit}
\end{figure}

Is it time to write off probe storage completely and learn from what
has been accomplished?  Recently, progress in the increase in
areal densities has slowed down, both in electronically addressed
(solid-state) memories and in mechanically addressed storage
(Figure~\ref{fig:AngstromsPerBit}). Inevitably, however, the issue of thermal
stability in hard-disk storage will force a transition to other
storage principles. This transition will have to occur in the
next one or two decades. Currently, probe storage remains the only
potentially viable route to achieving densities beyond those of the hard-disk.

\subsection{Paper outline}
This review is an updated and expanded version of parts of the book chapter `Probe Storage'~\cite{ProbeStorageChap2011} and is targeting the nanoscale aspects of a probe-storage system. It is divided into three sections: probe and medium technologies (section~\ref{sec:probe}), positioning systems
(section~\ref{sct:positioning}), and probe arrays and parallel readout
(section~\ref{sec:probe-array}). The probe and medium technologies section is split into subsections according to the different physical methods of storing data. In each such subsection, first the type of storage is introduced, then the write and read processes and the storage media are described. Finally, endurance is discussed, as it remains a key issue in further maturing probe technologies.

The positioning systems section is devoted to the positioning of the storage medium relative to the probe array. Different types of actuators are compared in terms of their suitability for a probe-storage system. The probe arrays and parallel readout section discusses the challenges in scaling up from a single probe as used in scanning probe microscopy to 2D~probe arrays as required for probe storage. 

%#################################################################
%%% Local Variables:
%%% mode: latex
%%% TeX-master: "PrStReview"
%%% End:

%#################################################################
\section{Probe and medium technologies}
\label{sec:probe}
In this section, we discuss the various principles of storing and
reading data. We can distinguish a number of categories, each with
its own physical parameter that is locally modified to store data:
(1) topographic storage uses a topographical change, (2) phase-change
storage in chalcogenide materials (e.g., {G}e$_2${S}b$_2${T}e$_5$) uses
a change in the phase of the material from amorphous to crystalline
and vice versa, (3) phase-change storage on non-chalcogenide materials
uses two distinct material phases, (4) magnetic storage uses
magnetization, (5) ferroelectric storage uses electrical polarization,
(6) atomic and molecular storage uses the relative orientation of
atoms. Especially this last technology has a strong potential for future work on probe storage. 

%%%%%%%%%%%%%%%%%%%%%%%%%%%%%%%%%%%%%%%
%__________SUBSECTION_________________%
\subsection{Topographic storage}

%\subsubsection{Introduction}
The most mature probe storage technology is topographic storage, mainly developed by IBM in a project termed `millipede'~\cite{Vettiger2000, Knoll2006}. A thermomechanical process creates a topographical change in a polymer medium. The change is, in the most straightforward implementation, an indentation that represents a 1. The absence of the indentation is used both as a spacer between neighboring 1s and for denoting a 0.

\subsubsection{Data writing}
\begin{figure}
  \centering
  \includegraphics[width=\figurewidth]{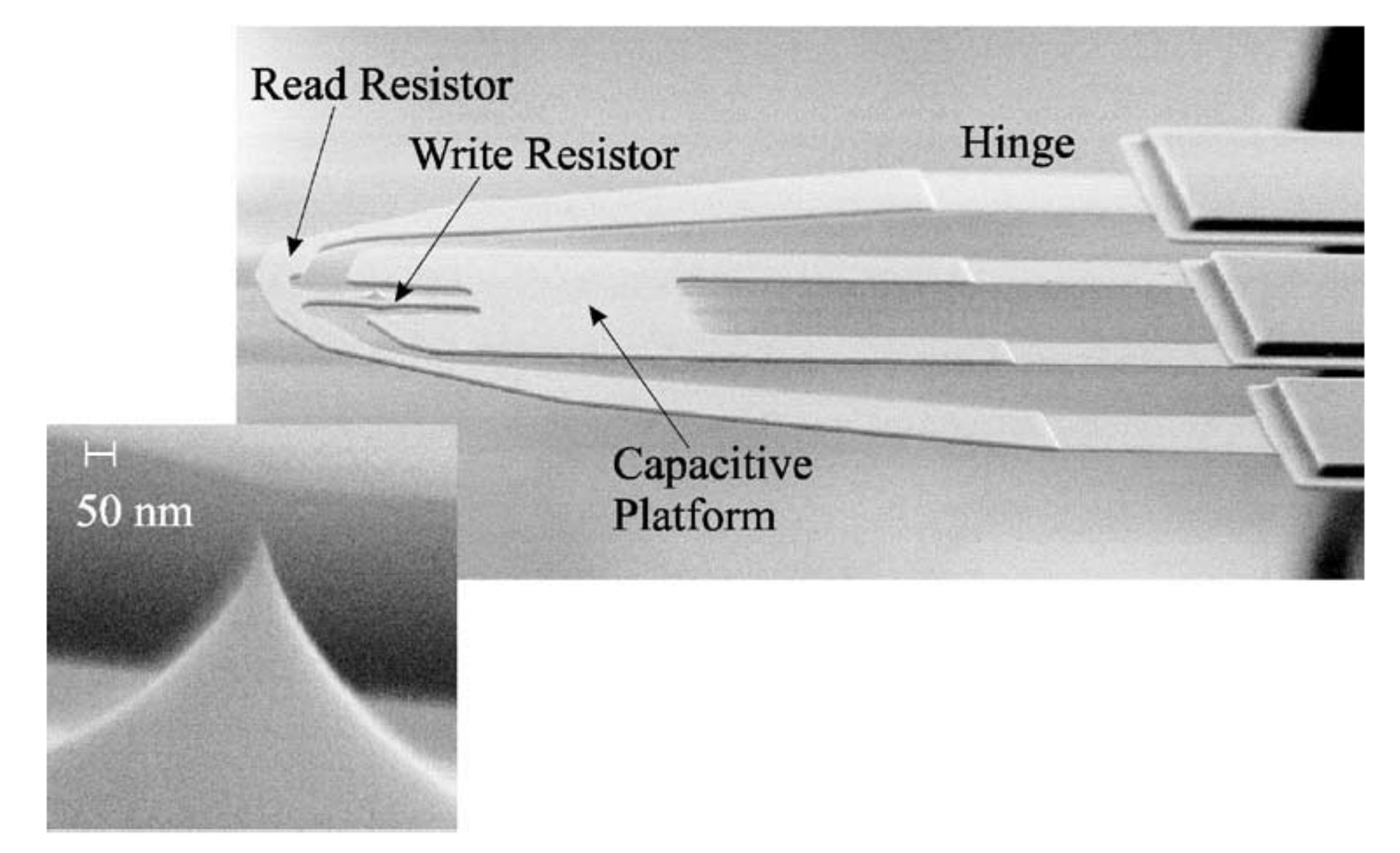}
  \caption{Scanning-Electron Microscopy (SEM) image of the three-terminal thermomechanical probe that was used to perform the read/write demonstration at \SI{641}{Gb~in^{-2}}. During read operation, the read resistor is heated to approximately \SI{250}{\celsius}; during write operation, the write resistor is heated to approximately \SI{400}{\celsius}. The inset shows an enlarged view of the sharp silicon tip located on the write resistor. Reproduced with permission from~\onlinecite{Pozidis2004}.}
  \label{fig:Pozidis2004}
\end{figure}

Writing is accomplished by heating the tip of the probe and applying an electrostatic force to the body of the cantilever, thereby pulling the tip into the medium.  The tip is heated by means of a localized heater at its base, see Figure~\ref{fig:Pozidis2004}. The heater consists of a low-doped resistive region of silicon that acts as heating element. This writing process has been demonstrated to be capable of megahertz writing speeds at densities above \SI{1}{Tb~in^{-2}}~\cite{Cannara2008}.

The development of this thermomechanical write process in polymers
started with the early work of Mamin~\etal~\cite{Mamin1992}. They used
an external laser to supply the heat to the cantilever stylus and
achieved heating times of \SI{0.3}{\micro\second} and data rates of
\SI{100}{kb~\second^{-1}}. The integration of the heater in the
cantilever initially led to an increase of the heating time to
\SI{2.0}{\micro\second}. A later design by King~\etal resulted in a decrease in the heating time down to \SI{0.2}{\micro\second}~\cite{King2001}. This design was realized using a combination of conventional and e-beam lithography~\cite{Drechsler2003}. The cantilevers in this design were \SI{50}{\micm} long and only \SI{100}{nm} thick, yielding extremely low spring constants of \SI{0.01}{N~m^{-1}}. The size of the heater platform was reduced down to \SI{180}{nm}, resulting in time constants on the order of \SI{10}{\micro\second}. The writing energy was less than \SI{10}{nJ} per bit, mainly because of parasitic effects and an inappropriate measurement setup, so there is potential for improvement.
The storage density is defined by the medium properties, and, more importantly, also by the probe tip dimensions. Lantz~\etal tried to achieve higher densities by applying multiwalled carbon nanotube tips with a tip radius down to \SI{9}{nm}. The advantage of the carbon nanotube tips is that the tip radius does not increase by wear, instead the tip just shortens. Densities of up to \SI{250}{Gb~in^{-2}} were reached~\cite{Lantz2003}, which was disappointing because at that time densities up to \SI{1}{Tb~in^{-2}} could already be attained with ultrasharp silicon tips. However, the power efficiency was improved because of better heat transfer through the nanotube. Data could be written at heater temperatures of \SI{100}{K} lower than with comparable silicon tips.

\subsubsection{Data reading}
To read back the data, a second resistor is present in one of the side-arms of the three-legged cantilever design, see Figure~\ref{fig:Pozidis2004}. This resistor acts as temperature-dependent resistor, whereby an increasing temperature causes a higher resistance. The read resistor is heated, and the amount of cooling is accelerated by proximity to the medium. When the tip reaches an indentation, the medium is closer to the read resistor and thus the current that flows through the resistor will increase. The data is read back by monitoring this current. The platform is heated to about \SI{300}{\celsius}, well below the temperature needed for writing, and a sensitivity of \SI{1e-5}{nm^{-1}} is obtained~\cite{Durig2000}.

Thermal readout was investigated in more detail by King~\etal, who showed that the fraction of heat transferred through the tip/medium interface is very small and most of the heat flow passes across the cantilever-sample air gap~\cite{King2002}. This observation presented the possibility of heating a section of the cantilever to avoid reading with heated tips, which can cause unwanted erasures and increased medium wear. Simulations were performed to optimize the probe design. A shorter tip increased the sensitivity to \SI{4e-4}{nm^{-1}}~\cite{King2001}.

To guide and speed up the design of more sensitive probes and assist in the readout data analysis, D\"urig developed a closed-form analytical calculation for the response of the height sensor~\cite{Durig2005}. An optimized design by Rothuizen~\etal led to a bandwidth of several tens of kHz at powers on the order of \SI{1}{mW}~\cite{Rothuizen2009}. Later, by using feedback, the read speed of the optimized design could be increased from \SI{19}{kHz} to \SI{73}{kHz}~\cite{Sebastian2009}.

\subsubsection{Recording medium}
\label{sec:TMRecMedium}
The first polymer media used for thermomechanical storage were simply $1.2$-mm-thick PMMA (perspex) disks~\cite{Mamin1992}. Using a single cantilever heated by a laser through the PMMA disk, Mamin~\etal were able to write bits having a radius below \SI{100}{nm} and a depth of \SI{10}{nm}, enabling data densities of up to \SI{30}{Gb~in^{-2}}. In subsequent work, the bulk PMMA or polycarbonate (compact disk material) disks were replaced by silicon wafers on top of which a $40$-nm PMMA recording layer on top of a $70$-nm cross-linked hard-baked photoresist were deposited. This allowed small bit dimensions down to \SI{40}{nm}, and data densities of up to \SI{400}{Gb~in^{-2}} were shown~\cite{Binnig1999}.
In addition to PMMA, also other polymers were studied, such as
polystyrene and polysulfone~\cite{Vettiger2002}. A write model was
developed by D\"urig (Figure~18 in~\cite{Vettiger2002}).  From the
model, it became clear that a balance needs to be found between stability and wear resistance of the medium on the one hand, requiring highly cross-linked polymers~\cite{Gotsmann2006b}, and wear of the tip on the other hand, for which a soft medium is necessary. 

Based on this knowledge, a so-called Diels-Adler (DA) polymer was introduced~\cite{Gotsmann2006}. These DA polymers are in a highly cross-linked, high- molecular-weight state at low temperature, but dissociate at high temperature into short strains of low molecular weight. This reaction is thermally reversible: rather than a glass-transition temperature, these polymers have a dissociation
temperature. Below the transition temperature, the polymer is thermally stable and has a high wear resistance; above the transition temperature the polymer becomes easily deformable and is gentle on the tip. Using the new DA polymer, densities of up to \SI{1}{Tb~in^{-2}} were demonstrated~\cite{Gotsmann2006}. 

\begin{figure}
  \centering
  \includegraphics[width=\figurewidth]{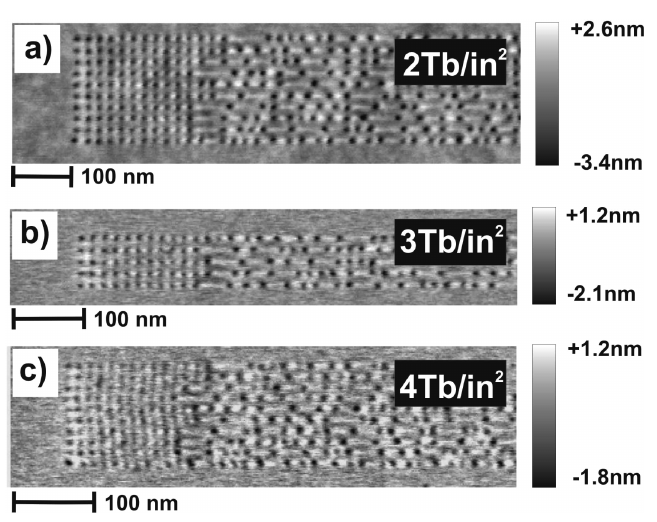}
  \caption{AFM images of a random pattern of indentations recorded in a PEAK polymer. A density of \SI{2}{Tb~in^{-2}} was obtained on a normal spin-coated sample. Densities up to \SI{4}{Tb~in^{-2}} were achieved on a templated sample. Reproduced with permission from~\onlinecite{Wiesmann2009}.}
  \label{fig:Wiesmann2009}
\end{figure}

The work was continued with a polyaryletherketone (PAEK) polymer, which incorporates diresorcinol units in the backbone for control of the glass-transition temperature and phenyl-ethynyl groups in the backbone and as endgroups for cross-linking functionality~\cite{Wiesmann2009}. As with the DA polymer, this polymer is highly crosslinked to suppress media wear during reading and to enable repeated erasing. In contrast to the DA polymer, however, it has a conventional, but very low, glass-transition temperature of less than \SI{160}{\celsius} in the cross-linked state, enabling indentation on a microsecond time scale using heater temperatures of less than \SI{500}{\celsius}. It exhibits exceptional thermal stability up to \SI{450}{\celsius}, which is crucial for minimizing thermal degradation during indentation with a hot tip. Using this polymer, densities of up to \SI{4}{Tb~in^{-2}} have been achieved on ultra-flat polymers made by templating the polymer on a cleaved mica surface~\cite{Pires2009, Knoll2010b}, see Figure~\ref{fig:Wiesmann2009}. Modeling shows that in this type of polymer media the density is limited to \SI{9}{Tb~in^{-2}}~\cite{Wiesmann2009}. 
The polymer crosslink density and topology have been optimized, leading to experimental demonstrations of \SI{1}{Tb~in^{-2}} density and $10^8$ write cylces using the same tip~\cite{Gotsmann2010}. The optimized polymer was shown to endure $10^3$ erase cycles and $3^.10^8$ read cycles and featured an extrapolated 10 year retention at \SI{85}{\celsius}.
Another approach is to introduce an ultrathin elastic coating to optimize the mechanical stability of the underlying polymer film~\cite{Kaule2013}. A plasma-polymerized norbornene layer physically separates the plastically yielding material from the material in contact with the tip. Both tip and medium wear are reduced, while retaining retention of the indentations written.

Apart from IBM, also others have been investigating polymer media. Kim~\etal from LG demonstrated bit diameters of \SI{40}{nm} in PMMA films~\cite{Kim2005}. Bao~\etal of the Chinese Academy of Sciences investigated friction of tips with varying diameters on PMMA, and concluded that blunt tips can be used to determine the glass-transition temperature, whereas $30$-nm-diameter tips can be used to detect local ($\beta$) transitions~\cite{Bao2008}.

\subsubsection{Endurance}
Endurance poses one of the largest problems for a probe storage system. Much work has been done to mitigate tip wear, which is the main issue that needs to be overcome to obtain a reasonable device life-time. To get a feel for the extent of the problem of tip wear, a description of the wear issue in thermomechanical storage is taken from~\cite{Lantz2009}. Consider a system operating at \SI{1}{Tb~in^{-2}} and a data-rate of \SI{100}{kb.s^{-1}} per probe. With the data storage industry life-time standard of 10~years and continuous operation of the device, each probe slides a distance of 10 to \SI{100}{km}. This translates into a maximum tolerable wear rate on the order of one atom per 10~m sliding distance in order to maintain the \SI{1}{Tb~in^{-2}} density. When operated in normal contact mode on a polymer medium, a silicon tip loaded at \SI{5}{nN} wears down in \SI{750}{m}, sliding to a bluntness that corresponds to data densities of \SI{100}{Gb~in^{-2}}.

A first estimate of the tip-sample force threshold at which wear starts to become an issue was given by Mamin~\etal~\cite{Mamin1995}. A load force of \SI{100}{nN} is mentioned to maintain reliable operation for the relatively large-sized indentations (\SI{100}{nm}) described in this early work. Such a force is detrimental for any reported probe-medium combination when densities above \SI{1}{Tb~in^{-2}} are targeted. In a more exhaustive study on wear by Mamin~\etal~\cite{Mamin1999}, a bit diameter of \SI{200}{nm} is shown to be maintained throughout a tip travel length of \SI{16}{km}. Although the tip travel length is sufficient for a probe storage device, the bit diameter is far from competitive. 

Several ways to reduce tip wear in thermomechanical recording have been proposed and demonstrated. Three important measures are discussed here.
 
A first measure to reduce the tip wear is softening of the medium, e.g., by the inclusion of a photo-resist layer of \SI{70}{nm} between the silicon substrate and the storage medium (PMMA)~\cite{Binnig1999}. Various other measures to reduce tip wear from the medium side have been investigated, see Section~\ref{sec:TMRecMedium} for details.

Hardening of the probe is a second way of mitigating tip wear. Coating the tip with a hard material or molding a tip generally leads to larger tip radii. The wear resistance of probes was reduced using diamond-like carbon tips to bring down wear to only one atom per every micrometer of tip sliding~\cite{Bhaskaran2010}. A further improvement was achieved by Lantz~\etal using SiC tips~\cite{Lantz2012}. For thermomechanical storage, silicon is therefore preferred~\cite{Lantz2009}.

A third way to reduce the tip wear is by actuation of the tip with a periodic force at frequencies at or above the natural resonant frequency of the cantilever. It is known from AFM that the intermittent-contact mode of operation reduces tip wear, and this is one of the foremost reasons that intermittent contact is preferred over contact mode in many microscopy environments. Application of the intermittent-contact mode for probe storage is not very straightforward. There are many requirements on the probes, and some of them conflict with the requirements for intermittent contact, e.g., the high cantilever stiffness required for intermittent-contact AFM conflicts with the feeble cantilever used in thermomechanical storage to allow easy electrostatic actuation. The speed of intermittent-contact modes is also reported to be insufficient for probe storage~\cite{Lantz2009}. In Ref.~\cite{Sahoo2008}, a solution is presented that uses amplitude modulation of the cantilever through electrostatic actuation despite a high nonlinearity in the cantilever response. The authors show successful read and write operation at \SI{1}{Tb~in^{-2}} densities after having scanned \SI{140}{m}. A second solution is to slightly modulate the force on the tip-sample contact. Lantz~\etal~\cite{Lantz2009} showed that by application of a \SI{500}{kHz} sinusoidal voltage between the cantilever and the sample substrate, tip wear over a sliding distance of \SI{750}{m} is reduced to below the detection limit of the setup used. Knoll~\etal~\cite{Knoll2010} use a similar actuation voltage between sample and tip to achieve a so-called dithering mode that is shown to effectively prevent ripples on a soft polymer medium. The authors attribute the absence of the otherwise present ripples to the elimination of shear-type forces by dithering at high frequencies.

%%%%%%%%%%%%%%%%%%%%%%%%%%%%%%%%%%%%%%%
%__________SUBSECTION_________________%
\subsection{Phase-change storage on chalcogenide media}

%\subsubsection{Introduction}
Phase-change storage is well-known from optical disks, for which laser light is used to modify phase-change materials, such as Ge$_\text{2}$Sb$_\text{2}$Te$_\text{5}$, to store information. Data storage is performed by locally changing an amorphous region to a crystalline region and vice versa. This transition is accompanied not only by a significant change in reflectivity, which is exploited for optical disks, but there is also a change in resistivity over several orders of magnitude. Major work has been done on probe recording on phase-change media at Matsushita~\cite{Tohda1995, Kado1997}, Hokkaido University~\cite{Gotoh2004}, CEA Grenoble~\cite{Bichet2004,Gidon2004}, the University of Exeter~\cite{Wright2006, Wright2003, Aziz2005, Wright2008, Wright2010, Wang2011, Wang2014, Wang2014a}, and Hewlett-Packard~\cite{Naberhuis2002}.

\subsubsection{Data writing}
Phase-change recording in probe storage uses an electrical current to induce the heat required. A conductive probe passes a current through the storage medium. The current locally heats the medium and, at sufficiently high temperatures, a transition from the amorphous to the crystalline phase is induced. This write process is self-focusing, resulting in bit densities  greater than \SI{1}{Tb~in^{-2}}~\cite{Gidon2004}. The power consumption for the writing process is low with respect to other technologies (smaller than \SI{100}{pJ} per  bit written~\cite{Satoh2006}). This is because only the bit volume, and not the entire tip volume, is heated.
There are, however, alternative strategies in which the tip itself is heated. Lee~\etal used a resistive heater to increase the tip temperature and write crystalline bits~\cite{Lee2002a}. Hamman~\etal achieved an impressive density of \SI{3.3}{Tb~in^{-2}} by heating the AFM probe with a pulsed laser diode~\cite{Hamann2006}.
The authors anticipate write speeds of \SI{50}{Mb~s^{-1}} for one probe when using a spinning disk to position the medium (as in hard-disk drives) and a nanoheater instead of the pulsed laser diode. Rewriteability is demonstrated by erasing part of the written data using a focused laser diode. The dynamics of the AFM tip is too slow to realize the fast thermodynamics needed for amorphization. In general, amorphizing phase-change materials with a probe is very challenging~\cite{Bhaskaran2009c}.
Phase-change storage offers the possibility for more advanced write strategies, such as multi-level recording~\cite{Burr2010} and mark-length encoding~\cite{Wright2010}. The latter holds promise to increase user densities by at least \SI{50}{\%} and potentially as much as \SI{100}{\%}.

At the Tohoku University, Lee~\etal used dedicated heater tips to write bits into AgInSbTe films~\cite{Lee2002a}. Readout was achieved by measuring the local conductance of the medium.

\subsubsection{Data reading}
The most common method of data read back is to measure the conductance of the medium by applying a low potential on the probe and monitoring the current. If the probe is in direct contact with the medium, one essentially performs conductive AFM~\cite{Bichet2004}. Also non-contact modes exist that rely on changes in either the field-emitter currents~\cite{Naberhuis2002} or the tip-sample capacitance by Kelvin probe force microscopy~\cite{Nishimura2002}.
The difference in material density between the amorphous and the crystalline phase can also be exploited. The crystalline phase has a higher density, causing a bit written in an amorphous background to appear as a valley that is several angstroms deep~\cite{Bichet2004, Gidon2004}. The topographic map of the surface can be obtained by standard tapping-mode AFM~\cite{Hamann2006}.

\subsubsection{Recording medium}
Phase-change recording media have mainly been researched for storage on optical disks and currently form an active field of research for non-volatile memory applications. 
Thorough overviews of solid-state phase-change memory are given in~\cite{Burr2010} and in~\cite{Wong2010}. A map for phase-change recording materials has been developed~\cite{Lencer2008} and their properties have been reviewed in detail~\cite{Wuttig2007}.

\subsubsection{Endurance}
Tip wear is quite a severe issue because not only the tip sharpness has to be maintained, but also the tip's ability to conduct. Tips for phase-change recording have been successfully made more wear-resistant by changing the fabrication material. The deposition of platinum on a silicon tip and subsequent annealing create a hard layer of platinum silicide~\cite{Bhaskaran2009a}. An ingenious way to strengthen the tip is encapsulation of the conductive platinum silicide tip with a relatively large layer of silicon oxide. The pressure on the tip apex is now decreased because of the increase of the tip area. The resolution of storage is, however, still determined by the small conductive core~\cite{Bhaskaran2009,Bhaskaran2009b}. Such a design leads to more stringent demands on the medium side, as the larger tip apex will typically generate larger forces at the tip-sample interface, thereby potentially wearing down the medium.
Force-modulation schemes are shown to be beneficial for the endurance of conductive probes. Moreover, it has been demonstrated that when using force modulation lower, load forces are needed to obtain a good electrical tip-medium contact~\cite{Koelmans2010a, Koelmans2011}.
Given the severe limitations in writing amorphous regions in phase-change materials, the prospects for probe-based phase-change storage are dim. 
The application of probes on phase-change media is, however, very useful to study material properties at the nanoscale, and much of the insight gained for probe-storage purposes can be applied in this context.

%%%%%%%%%%%%%%%%%%%%%%%%%%%%%%%%%%%%%%%
%__________SUBSECTION_________________%
\subsection{Phase-change storage on non-chalcogenide media}
Although most work is done on chalcogenide materials, a number of alternatives exist and are briefly summarized here. 

Instead of phase-change media based on inorganic compounds, such as GeSbTe alloys, one can use polymers that become conductive upon application of a voltage~\cite{Takimoto1997}. The change in conductivity can be due to a change in the phase, caused for instance by polymerization~\cite{Shi2000}, or due to electrochemical reactions~\cite{esashi2005,yoshida2005,yoshida2007}. The latter method is especially interesting as it is reversible. The exact nature of the reaction is unknown; it could be either an oxidation-reduction or protonation-deprotonation reaction. Polymer media are softer than alloys, and tip wear is expected to be less of an issue~\cite{Yoshida2013}. Rewritability, however, could be a problem because the polymer tends to polymerize. Rather than heating the phase-change material by passing a current from tip to sample, one can use heated tips. 

Phase-change storage without the use of heat has been demonstrated at \SI{1}{Tb~in^{-2}}~\cite{Jo2009}. The researchers from LG Electronics and Pohang University in Korea use the tip to apply pressure alone, causing microphase transitions of the polystyrene-block-poly (n-pentyl methacrylate) of a block copolymer.

%%%%%%%%%%%%%%%%%%%%%%%%%%%%%%%%%%%%%%%
%__________SUBSECTION_________________%
\subsection{Magnetic storage}

Magnetic recording is one of the oldest data-storage technologies, and
many researchers have attempted to write on magnetic media with
probes. The reasons are simple: Magnetic recording materials
are readily available, and in recording labs, the Magnetic Force
Microscope (MFM) is a standard instrument.

\subsubsection{Data writing}
The stray field of MFM probes is relatively low,
which limits the maximum achievable density to about
\SI{200}{Gb in^{-2}}~\cite{Mironov2009}. Moreover, dots can only be
magnetized in the direction of the tip magnetization. Under these
constraints, however, MFM writing has been beautifully demonstrated by
Mironov~\etal~\cite{Mironov2009a}.  Writing experiments have
been performed in a vacuum MFM on Co/Pt multi-layered dots with
perpendicular anisotropy. Two samples were investigated, one with a
dot diameter of \SI{200}{nm} and a periodicity of \SI{500}{nm}, and
the other with a diameter of \SI{35}{nm} and a spacing of \SI{120}{nm}. The
large dots could be written by moving the MFM tip in contact over the
medium. The $35$-nm dots could be written by merely touching the
dots with the MFM tip.

To write onto modern recording media at higher density, some
type of assist will be necessary. There are essentially two methods:
applying a uniform external background field or applying heat.

% External field
An external field can easily be applied by means of a small coil
mounted below the medium. As early as 1991, Ohkubo~\etal{}
used permalloy tips on a CoCr film for perpendicular
recording~\cite{Ohkubo1991,Ohkubo1993,Ohkubo1993a}. By applying the
field in opposite directions, the magnetization of the tip can be
reversed, and higher bit densities can be obtained by partially
overwriting previously written bits.  Bit sizes down to \SI{150}{nm} could
be obtained~\cite{Ohkubo1995}, and overwriting data was
possible. Similar bit sizes were obtained by
Manalis~\cite{Manalis1995} using a CoCr alloy and CoCr- or NiFe-coated tips.

The bit sizes are relatively large, limiting data densities to
somewhere on the order of \SI{30}{Gb in^{-2}}. This is either due to the
media used or the limited resolution of the MFM tip. Detailed
analyses by El-Sayed showed, however, that densities up to
\SI{1.2}{Tb in^{-2}} should be possible with a rather conventional $30$-nm tip
radius~\cite{El-Sayed2003,El-Sayed2004}.

Onoue~\etal{} showed that care must be taken when applying high voltages
to coils below the medium. If the medium is not 
grounded, a large capacitive charging current will flow from the tip
into the sample, unintentionally heating the medium~\cite{Onoue2008}
and resulting in relatively large bits. Without grounding, no bits
could be written because of the high switching field distribution in the
Co/Pt multilayer used.

To increase the tip's stray field, Kappenberger~\etal{} produced Co rod-like
tips of \SI{100}{nm} diameter by means of electrodeposition
inside a porous alumina membrane~\cite{Kappenberger2009}. The tip apex
was tailored down to a \SI{25}{nm} diameter by means of focused ion-beam
etching. Using this tip and a UHV MFM, they performed write
experiments in 60-nm-diameter dots deposited on nano-spheres
spaced by \SI{100}{nm} (which would be equivalent to
\SI{60}{Gb/in^{-2}}). Even though the tip's stray field was expected to be
large, a background field of \SI{405}{mT} still had to be applied to
reverse the magnetization.

\begin{figure}
  \centering
  \includegraphics[width=8cm]{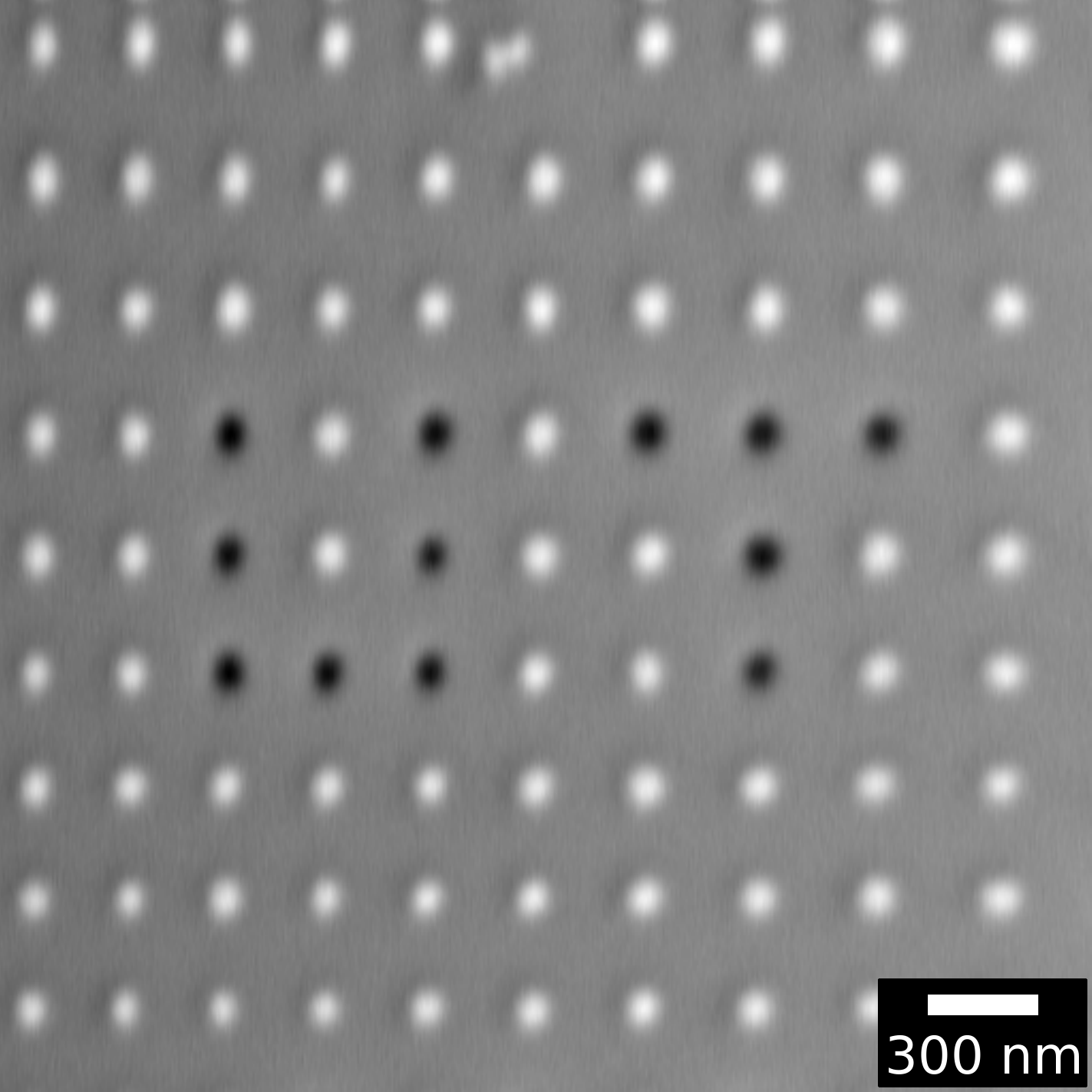}
  \caption{Magnetic force microscopy image of the letters UT (of
    University of Twente) written in an array of Co/Pt multilayered
    islands, patterned by laser interference lithography and ion-beam
    etching.}
  \label{fig:MFM_UT}
\end{figure}

Another example of tip-field-induced
writing in bit-patterned media is shown in
Figure~\ref{fig:MFM_UT}. The medium is based on a Co/Pt multilayered
film with perpendicular anisotropy, patterned by Laser Interference
Lithography~\cite{DeVries2013}. Because the switching field distribution
of the magnetic islands is larger than the tip's stray field, several
tricks had to be applied. A uniform external magnetic field was
applied under an angle of \SI{45}{\degree} to achieve the lowest
possible switching field distribution. First, a field sweep is
performed to determine the order of switching of the islands. The
pattern is then written in such a way that the islands that switch at
the highest field are written first. For this, the external magnetic
field is increased to a value just below the switching field of the island. The
combined field of the external uniform field and the tip's stray field
then reverses only this island.  To maximize the effect of the tip's stray field,
the tip is lowered several times to the island in closely spaced
points, while the MFM slow-scan direction is slightly
modulated. This ensures that uncertainties in the tip position are
compensated and that occasional domains are pushed out of the
island~\cite{Mironov2009}. Once the strongest island has been written, the
external field is slightly reduced and the procedure repeated until
the last island has been reversed.

% Heat assist
% Using demag field.
Rather than applying background fields and incurring the risk of erasing
previous information, one can locally heat the medium to reduce its switching
field. This is a method also suggested for future hard-disk systems
with extremely high-anisotropy media~\cite{Ruigrok2000,Seigler2008}.

In contrast to hard-disk recording, it is surprisingly easy to deliver heat to the medium in probe storage. The most straightforward method
for heating is to pass a current from the tip to the medium. Watanuki~\etal~\cite{Watanuki1991} used an STM tip made
from an amorphous magnetic material around which a coil was wound.
The tip-sample distance was controlled by the tunneling current. Bit sizes
on the order of \SI{800}{nm} were achieved.

For testing purposes, one does not even have to use a magnetic tip or
apply a background field: When starting from a perpendicularly
magnetized film, the demagnetization field of the surrounding film
will reverse the magnetization in the heated area. This procedure
allows write-once experiments. Hosaka~\etal{} experimented with
writing bits into magnetic films by passing a current from an STM tip
into a Co/Pt multilayer with perpendicular anisotropy. The minimum domain size, observed by optical microscopy, was 250
nm~\cite{Hosaka1995,Nakamura1995}, but smaller domains might have been
present. The experiment was repeated by Zhang~\etal{} but now
the bits were imaged by 
MFM~\cite{Zhang2007,Zhang2006,Zhang2006a,Zhang2006b,Zhang2006c,Zhang2006d,Zhang2004}. 
Bits sizes down to \SI{170}{nm} were observed in Co/Pt multi-layers and
weakly coupled CoNi/Pt granular media by Zhong~\cite{Zhong2007}.

\begin{figure}
  \centering
  \includegraphics[width=6cm]{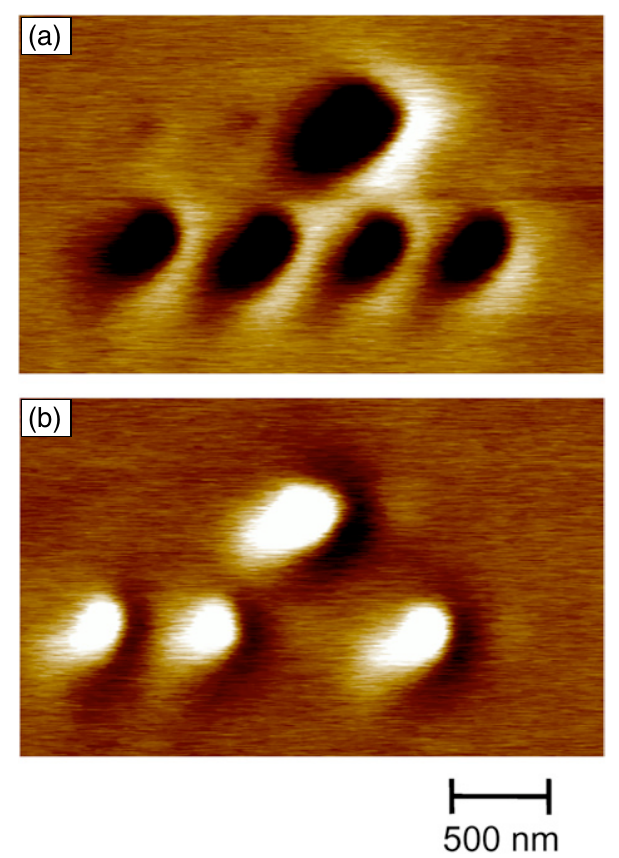}
  \caption{Magnetic force microscopy images demonstrating the erasure
    of individual bits in a Co/Pt multilayer. From ref.~\onlinecite{Onoue2008}, \copyright \ IOP Publishing. Reproduced by permission of IOP Publishing. All rights reserved.}
  \label{fig:Onoue2008}
\end{figure}

The large disadvantage of using STM~tips is that direct imaging of
written bits is only possible by spin-polarized tunneling, which is a
difficult technique. Using MFM~tips, imaging can be done
immediately after writing. Hosaka~\etal{} used an MFM~tip in
field-emission mode~\cite{Hosaka1995} and wrote bits as small as
\SI{60 x 250}{nm}. Onoue~\etal{} combined this method with
applying a pulsed background field~\cite{Onoue2004,Onoue2005}, so that
bits could also be erased~\cite{Onoue2008} (Figure~\ref{fig:Onoue2008}). The
minimum bit size obtained was \SI{80}{nm}, which is close to the bubble
collapse diameter for the Co/Pt films used in these
experiments~\cite{Onoue2008}.

% Heated tips
Rather than using currents, one can also use heated tips, similar to
those used on the polymer media described above. Algre~\etal{} proposed to write by means of a heated AFM
tip~\cite{Algre2005a}. They start from a Co/Pt multilayer patterned
medium prepared by sputtering on pillars of \SI{90}{nm} diameter, spaced
\SI{100}{nm} apart.  The pillars were etched into nano-porous silicon to achieve
good thermal insulation.  The authors show that these media are suitable
for heat-assisted magnetic probe recording. The readout method is, however, not
clear, and the authors do not demonstrate actual recording experiments.

\subsubsection{Data reading}
% Readback, MFM or MRM
Readback of magnetic information can be done by techniques based on MFM.
MFM being a non-contact mode, however, the
operation is complex. More importantly, the resolution is
determined by the tip-sample distance.  This distance has to be smaller
than the bit period, i.e., on the order of \SI{10}{nm} or less. At this
distance, non-magnetic tip-sample interactions become important,
leading to undesired topographic cross-talk. A straightforward
solution would be to use a Hall sensor integrated on a magnetic
tip~\cite{Sarajlic2010,Petit2004,Oral1996,Gregusova2009, Hatakeyama2014}.
For better signal-to-noise ratios, integration of a
magneto-resistive sensor at the end of the probe, similarly as in hard-disk
recording, is preferred.  An initial step in this direction was
taken by Craus~\etal{} using scanning magnetoresistance
microscopy~\cite{Craus2005}. The magnetic layer in the probe can be used as a 
flux-focusing structure, so that the same probe can be used for
writing. A more advanced spin-valve sensor was
integrated on a cantilever by Takezaki~\etal~\cite{Takezaki2006}. The
resolution was limited to about \SI{1}{\micm}, which is insufficient
for probe-based data storage. Magnetic field sensors integrated in
modern hard disks are, however, capable of resolutions far below \SI{20}{nm},
so in principle, the technique could be applied.

\subsubsection{Recording medium}

The data density in magnetic recording media is primarily determined
by the thermal stability of the written information. The energy
density in these media is relatively low compared with other
media (on the order of \SI{}{MJ m^{-3}}), so the problem was first
recognized in the field of magnetic storage~\cite{Charap1997}. In
essence, the energy barrier between the two information states -- in
magnetic recording the states are two opposing magnetization directions
 -- should be much higher than the thermal energy. At room
temperature, this means that the energy barriers should be higher than
\SI{40}{k_BT}, or approximately \SI{1}{eV}, which is a convenient
value to remember. The energy barrier is determined by the energy
density in the magnet and the volume of the bit. With increasing
density, the bit volume decreases, so the energy density in the
material should increase. The highest energy density known to us
(\SI{17}{MJ m^{-3}}) is found in SmCo alloys~\cite{Lectard1994}. At
this value, the minimum magnetic volume for stable storage is
approximately \SI{2.1}{nm^3}. Developments in hard-disk storage are
targeting storing one bit of information in this tiny volume, for
instance by using bit-patterned media~\cite{Lodder2004,Terris2007} or
by using aggressive coding techniques to store a bit in one or two
magnetic grains~\cite{Wood2009}. If successful, the maximum user
storage-density (this means after coding) imaginable with magnetic
recording will be on the order of a few \SI{10}{Tb
in^{-2}}~\cite{Kryder2008}, which is indicated at the
super-paramagnetic limit in Figure~\ref{fig:AngstromsPerBit}. Magnetic
probe-storage demonstrations, however, have been limited to a density
on the order of a few \SI{10}{Gb/in^2}.  In view of the fact that
values of up to \SI{4}{Tb in^{-2}} have already been demonstrated in
polymer-based media~\cite{Knoll2010b}, the question arises whether
magnetic probe-based storage should be pursued any further.

%\bibliography{paperbase}
%%% Local Variables: 
%%% mode: latex
%%% TeX-master: "PrStReview"
%%% End: 

%%%%%%%%%%%%%%%%%%%%%%%%%%%%%%%%%%%%%%%
%__________SUBSECTION_________________%
\subsection{Ferroelectric storage}
%\subsubsection{Introduction}
The electric counterpart of magnetic recording, ferroelectric storage, has been investigated for decades. In ferroelectric media, the domain walls are extremely thin, indicating a very high anisotropy. A promising piezoelectric material, such as PZT, has a typical coercive electric field of $10$--\SI{30}{MV~m^{-1}}~\cite{Pertsev2003} and a polarization of \SI{0.5}{C~m^{-2}}~\cite{Zybill2000}. The energy densities therefore appear to be on the order of 5-\SI{15}{MJ~m^{-3}}, which is a factor of two above the highest ever reported energy densities for magnetic materials. More important, however, is that the write head field is not material-limited, in contrast to the yoke in the magnetic recording head.

\subsubsection{Data writing}
Domain reversal is achieved by a conductive cantilever that is in contact with, or in close proximity to, the medium. It is reported that a voltage pulse as short as \SI{500}{ps} can successfully switch domains~\cite{Cho2006}. However, actual data rates realized are \SI{50}{kb~s^{-1}} per probe because of the low speed of the piezoelectric scanner used.

Franke~\etal at IFW Dresden~\cite{Franke1994} were the first to demonstrate the modification of ferroelectric domains by conductive AFM~probes. In their case, the probe was in contact with the surface, and writing was achieved simply by applying a tip-sample voltage of up to \SI{30}{V}.  Later, Maruyama~\etal at Hewlett-Packard in Japan obtained storage densities of up to \SI{1}{Tb~in^{-2}}~\cite{Hidaka1996, Maruyama1998}.
% Seagate
Rather than using probes, Zhao~\etal at Seagate realized a read/write head similar to hard-disk heads, where bits are defined at the trailing edge of the head~\cite{Zhao2008}. Using this novel type of head, densities up to \SI{1}{Tb~in^{-2}} were demonstrated.

\subsubsection{Data reading}
It is not entirely clear which method of data readback currently offers the best performance. A very fast method offering MHz rates at domain dimensions on the order of \SI{10}{nm} was demonstrated by Seagate~\cite{Forrester2009}. However, reading is destructive, as in conventional FeRAM, which adds significant complexity to the storage system. A constant read voltage is applied to a conductive probe, causing reversely polarized domains to switch. When this happens, the surface screening charge will change polarity. The current required is supplied and measured by electrical circuitry connected to the probe.

Readout of the polarization state of ferroelectric domains is usually accomplished by piezo force microscopy (PFM). PFM monitors the response of the probe to a small AC tip-sample voltage at a frequency below the cantilever resonance~\cite{Hidaka1996}. The sample thickness varies with this frequency because of the piezoelectric effect, and with twice this frequency because of electrostriction. Note, however, that on application of an electric field, also the permittivity changes, which gives rise to second harmonics~\cite{Franke1994}.
 
Readout can also be performed in non-contact mode. In the early nineties, Saurenbach and Terris at IBM Research-Almaden induced and imaged charges in polymer disks -- with tungsten probes~\cite{Saurenbach1990,Saurenbach1992}. Imaging was done in non-contact mode by measuring the electric field generated by the polarization charges. Saurenbach measured in dynamic mode, monitoring the changes in the resonance frequency of the cantilever caused by changes in the force derivative.
% Tohoku, Pioneer
At Tohoku University, ferroelectric probe-storage research started in the same period with experiments on PZT by Lee~\etal~\cite{Lee2002a} and on LiTaO$_3$ by Cho~\etal~\cite{Cho2002}. A frequency-modulation technique was used for data readout. The method is based on the fact that the storage medium's capacitance changes slightly on reversal of the ferroelectric polarization because of the non-linear terms in the permittivity tensor. This minute change in capacitance causes tiny changes in the resonance conditions, which can, for instance, be detected by monitoring the cantilever vibration if the cantilever is excited with a fixed AC voltage, preferably using a lock-in technique. Another method reported the direct piezoelectric effect to build up charge on the tip as a result of the tip-sample load force~\cite{Kim2009}. The resulting current is proportional to the load force, leading to a trade-off with endurance, as tip wear increases with the load force.

\subsubsection{Recording medium}
The maximum densities achieved are \SI{10.1}{Tb~in^{-1}} on a LiTaO$_3$ single-crystal medium~\cite{Cho2005} and \SI{3.6}{Tb~in^{-1}} on an atomically smooth PZT medium~\cite{Tayebi2010}. The storage areas are \SI{40x50}{nm} and \SI{1x1}{\micro m}, respectively.

% KAIST, Samsung
In 2000, Shin~\etal at KAIST experimented with AFM data storage on sol-gel deposited PZT~\cite{Shin2000}. Dots, with diameters on the order of \SI{60}{nm} to \SI{100}{nm}, were written at \SI{14}{V}. Data was read back by measuring electric forces in either non-contact or contact mode. Data retention appeared to be a problem, either because free charges accumulated on the medium surface or polarization was lost. 
Later work in collaboration with Samsung revealed that the polycrystalline nature of sol-gel deposited PZT films~\cite{Kim2006a} limits the data density, similarly as in hard-disk storage, and the authors concluded that the grain size needs to be decreased.

\begin{figure}
  \centering
  \includegraphics[width=9cm]{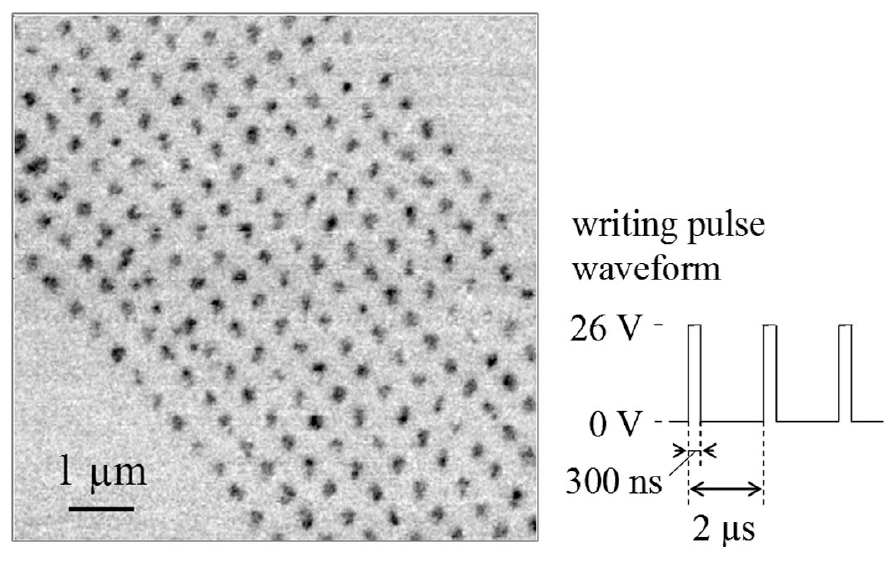}
  \caption{Scanning non-linear dielectric microscopy image of bits
    written in a LiTaO$_3$ film at a bit rate of \SI{1}{Mb~s^{-1}}. The bit spacing is
    \SI{206}{nm}. Reproduced with permission from~\onlinecite{Hiranaga2007a}.}
  \label{fig:Hiranaga2007a}
\end{figure}

Experiments at Tohoku University were continued on
LiTaO$_3$~\cite{Hiranaga2002, Cho2003, Hiranaga2003, Cho2004}, which
has superior stability.  As epitaxial films were used, pinning sites
are needed for thermal stability~\cite{Cho2003a}. By using thin
(\SI{35}{nm}) LiTaO$_3$ single-crystal films and a background field,
arrays of domains could be written at a density of
\SI{13}{Tb~in^{-2}}~\cite{Tanaka2008}. A realistic data storage
demonstration was given at \SI{1.5}{Tb~in^{-2}}. A raw bit-error rate
below $10^{-4}$ could be achieved at a density of \SI{258}{Gb~in^{-2}}
and at data rates of \SI{12}{kb~s^{-1}} for reading and
\SI{50}{kb~s^{-1}} for writing~\cite{Hiranaga2007a}, see Figure~\ref{fig:Hiranaga2007a}. Data retention was measured by investigating the readback signals at elevated temperatures, and an activation energy of \SI{0.8}{eV} at an attempt frequency of \SI{200}{kHz} was found, which is sufficient for a data retention of 10~years~\cite{Tanaka2008}. An overview of the work at Tohoku University until 2008 can be found in~\cite{Tanaka2008a}. A spin-off of this activity has started at Pioneer, mainly in the production of probes~\cite{Takahashi2007, Takahashi2006a, Takahashi2004, Takahashi2009}.

A later study investigated the concern about the bit stability when the ferroelectric domains get smaller than \SI{15}{nm}~\cite{Tayebi2010a}. The authors fully reversed domains through the entire ferroelectric film thickness, and thereby achieved stable domains with diameters down to \SI{4}{nm}.

The company Nanochip developed with Intel a conceptual prototype based on PZT media~\cite{Belov2009}. Adams~\etal report finding a non-destructive readout method that should bring their prototype much closer to commercialization~\cite{Adams2009}. However, in May 2009 Nanochip shut its doors and the attempt to commercialize the prototype was stopped.

\subsubsection{Endurance}
As is done in phase-change media, force modulation has also been applied in the readout of ferroelectric media to increase the endurance of the tip. A slide test where a platinum iridium tip slides for \SI{5}{km} at \SI{5}{mm~s^{-1}} was performed, where the authors claim that no loss in either the read or the write resolution occurred~\cite{Tayebi2010}. The total wear volume was estimated to be \SI{5.6e3}{nm^3}, which is impressively low. The result has been achieved at a load force of \SI{7.5}{nN} and with the application of force modulation at an amplitude of \SI{11}{nN}.

Another approach to increase the endurance of the tip makes use of a dielectricly-sheathed carbon nanotube probe that resembles a `nanopencil'~\cite{Tayebi2008}. These micrometer-long tips with constant diameter can sustain significant wear before the read and write resolution they provide decreases. Domain dots with radii as small as \SI{6.8}{nm} have been created.
Yet another approach is the use of hard HfB$_2$ tip-coating that can potentially extend the tip's endurance to beyond \SI{8}{km} of sliding~\cite{Tayebi2012}.

Tip wear can effectively be prevented by operating in non-contact mode; however, non-contact mode leads to lower data densities~\cite{Hiranaga2009}.

%%%%%%%%%%%%%%%%%%%%%%%%%%%%%%%%%%%%%%%
%__________SUBSECTION_________________%
\subsection{Atomic and molecular storage}
With ever shrinking bit dimensions, it is inevitable that mechanically addressed data storage will become impossible in continuous thin films, whether they are polymer-based, ferroelectric, magnetic or phase change. We will ultimately end up at the single molecule or atomic level. That this is not mere science fiction is elegantly proved in both molecular and atomic systems.

Cuberes, Schlitter and Gimzewski at IBM Research-Zurich demonstrated as early as 1996 that C$_{60}$ molecules can be manipulated and positioned on single-atomic Cu steps with an STM~\cite{Cuberes1996}. The experiments were performed at room temperature, and molecules remained stable during imaging. If the binding energy of the molecules is above 1~eV, this method could indeed be used for long-term data storage. Instead of fullerenes, which bind by Van der Waals forces, Nicolau~\etal  suggest to use ionic and chelation bonds between the molecules and the metal surface~\cite{Nicolau2004}. 

\begin{figure}
  \centering
  \includegraphics[width=9cm]{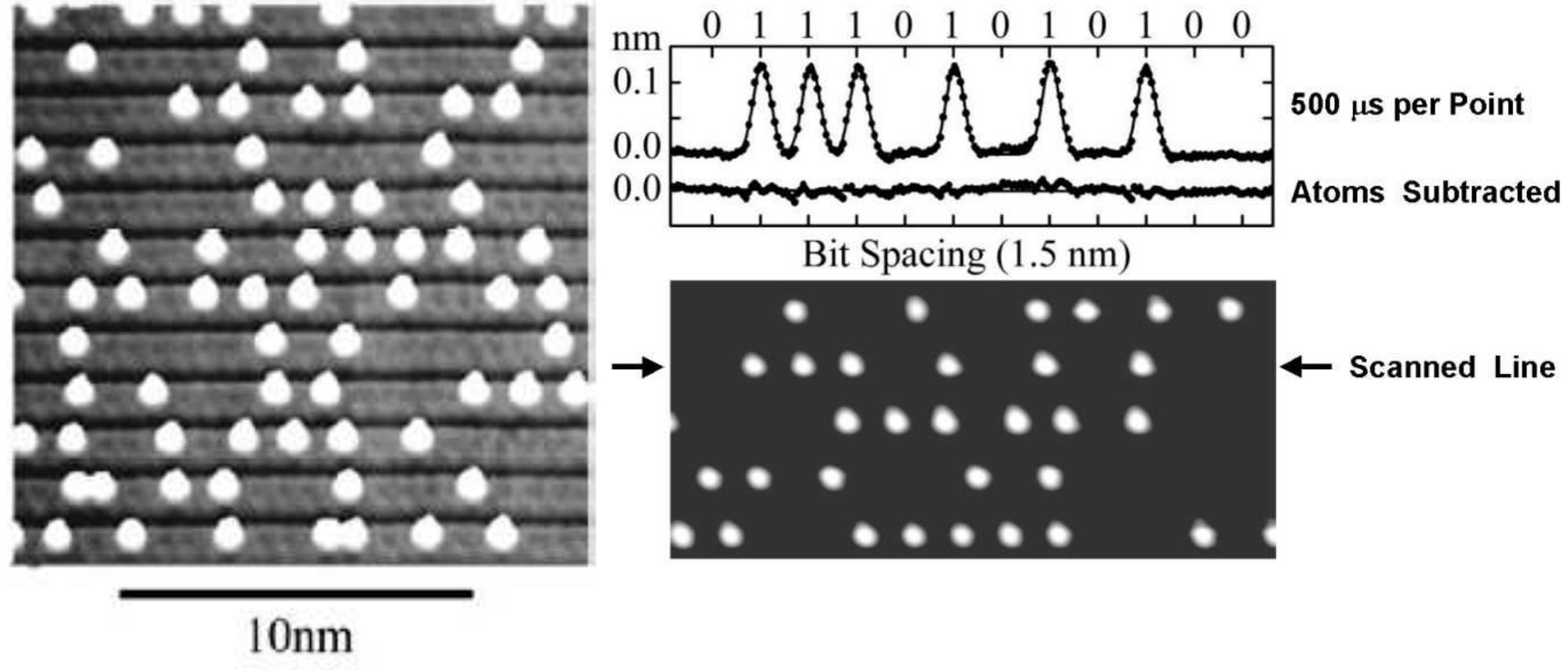}
  \caption{Atomic storage at room temperature: silicon atoms are positioned on top of a reconstructed silicon surface, leading to a density of \SI{250}{Tb~in^{-2}}. From ref.~\onlinecite{Bennewitz2002} \copyright \ IOP Publishing. Reproduced by permission of IOP Publishing. All rights reserved.}
  \label{fig:bennewitz}
\end{figure}

Storage of data in single atoms was demonstrated beautifully by Bennewitz~\etal in 2002~\cite{Bennewitz2002}, who deposited silicon atoms from an STM tip onto a 5$\times$2 reconstructed silicon-gold surface (Figure~\ref{fig:bennewitz}). Because of the nature of the reconstructed surface, every bit is stored into an area of 20 surface atoms, resulting in a density of \SI{250}{Tb~in^{-2}}. The method used by Bennewitz is a write-once technique, but one can also envision deposition of atoms from the gas phase, using, for instance, hydrogen~(H) or chlorine~(Cl)~\cite{Bauschlicher2001, Rosi2001}.

Even higher storage densities can be achieved if one does not use the position of molecules or atoms, but modifies their state. For molecules, one could use conformal changes, or change the charge state of single atoms~\cite{Repp2004}. Another option is to store data in the atomic spin~\cite{Loth2010}.

%%% Local Variables:
%%% mode: latex
%%% TeX-master: "PrStReview"
%%% End:

%#################################################################
\section{Positioning systems}
\label{sct:positioning}

A nanopositioner controls the position of an object with an accuracy on the order of nanometers. A probe data-storage system requires a miniature 2D nanopositioner, referred to as `the scanner', to move the storage medium relative to the probe array.
To be able to fully address the medium area, the displacement range must be equal to or larger than the distance between the probe tips in the probe array. This range, which is on the order of \SI{100}{\micm} (see Section~\ref{sec:probe-array}), must be combined with an accuracy on the order of a few nanometers.
The access time of a probe data-storage system is mainly determined by the positioning system.
The moving mass, the suspension spring stiffness, and the maximum actuator force determine this access time.
The mechanical rigidity of the scanner puts a lower bound on the moving mass. The spring suspension must be sufficiently stiff to prevent undesired resonances and to provide shock resistance and vibration rejection.
The actuator force should therefore be as large as possible, or at least on the order of millinewtons.

Several actuator types have been used for nanopositioner designs, such
as piezoelectric, electromagnetic, electrostatic (e.g., comb drive and
`inchworm'\footnote{``An inchworm moves by gripping a surface with its
  hind legs while retracting its body'', Hubbard et al.~\cite{Hubbard2006}.}),
electrothermal, and electrochemical actuators~\cite{Hubbard2006}.  The
size of a probe data-storage device is an important constraint,
severely restricting the space available for the nanopositioner.  The
actuators and scanner mechanics are commonly fabricated using
microelectromechanical systems (MEMS) technology.  The comparison of
MEMS actuators by Bell~\etal{} provides insight into which actuator
types are most appropriate for a probe data-storage
device~\cite{Bell2005}.  There are several suitable MEMS actuator
types, whose displacement range is on the order of \SI{100}{\micm}
with nanometer resolution and whose maximum force is on the order of
millinewtons. Electromagnetic, electrostatic (comb drive, dipole
surface drive, inchworm), thermal, and piezoelectric actuators all
seem promising candidates for use in a probe data-storage system, and
scanner designs have been published for all these actuator types
except for electrothermal actuators. A probable reason for the absence
of thermally actuated scanners for probe data-storage is the high
power required for fast thermal actuators.

Operating the scanner in closed-loop control is necessary to obtain nanometer positioning precision and adequate shock resistance~\cite{Devasia2007}. The position sensors should have a large dynamic range; namely, nanometer accuracy over a 100-\si{\micm} displacement range.
Suitable position sensors have a small footprint and are based on a varying thermal conductance~\cite{Lantz2005, Krijnen2011jmm, Zhu2011thermcap, Fowler2013, Mohammadi2014}, on a varying capacitance~\cite{Chu2003, Lee2009, Pang2009, Huang2010, Zhu2011thermcap}, on the position-dependent field of a permanent magnet~\cite{Kartik2012, Tuma2014GMR}, or on the piezoelectric effect~\cite{Messenger2009, Bazaei2014}.
Dedicated servo-pattern fields can enhance the positioning precision by providing a medium-derived position-error signal~\cite{Eleftheriou2003, Pantazi2007, Sebastian2008nanopos, Sebastian2012nanopos}.

%################################################################################

\subsection{Electrodynamic actuation}

Electromagnetic scanners use a coil to generate a magnetic field that leads to a force.
All electromagnetic actuators designed for probe storage reported in the literature use a permanent magnet and a coil. An electromagnetic comb-drive actuator without permanent magnet has been constructed~\cite{Schonhardt2008}, but has not been used in a scanner design yet. To distinguish electromagnetic actuators without and with permanent magnets, actuators with permanent magnets are referred to as electrodynamic actuators.

An advantage of electrodynamic scanners is the relatively straightforward linear actuator design (linear displacement-vs.-current curve), which simplifies controller design.
Another advantage for mobile probe storage is that an electrodynamic scanner can operate at the generally low voltage available because it is current driven. 
A disadvantage is that permanent magnets are needed. Assembling an electrodynamic scanner is therefore more complicated than assembling, for instance, an electrostatic comb-drive scanner.
It also means that it is difficult to make the scanner very thin.
The energy consumption of electromagnetic scanners in general is relatively large because of the large currents required and the series resistance of the coils~\cite{Engelen2013energy}.

\begin{figure}
  \centering
  \includegraphics[width=6cm]{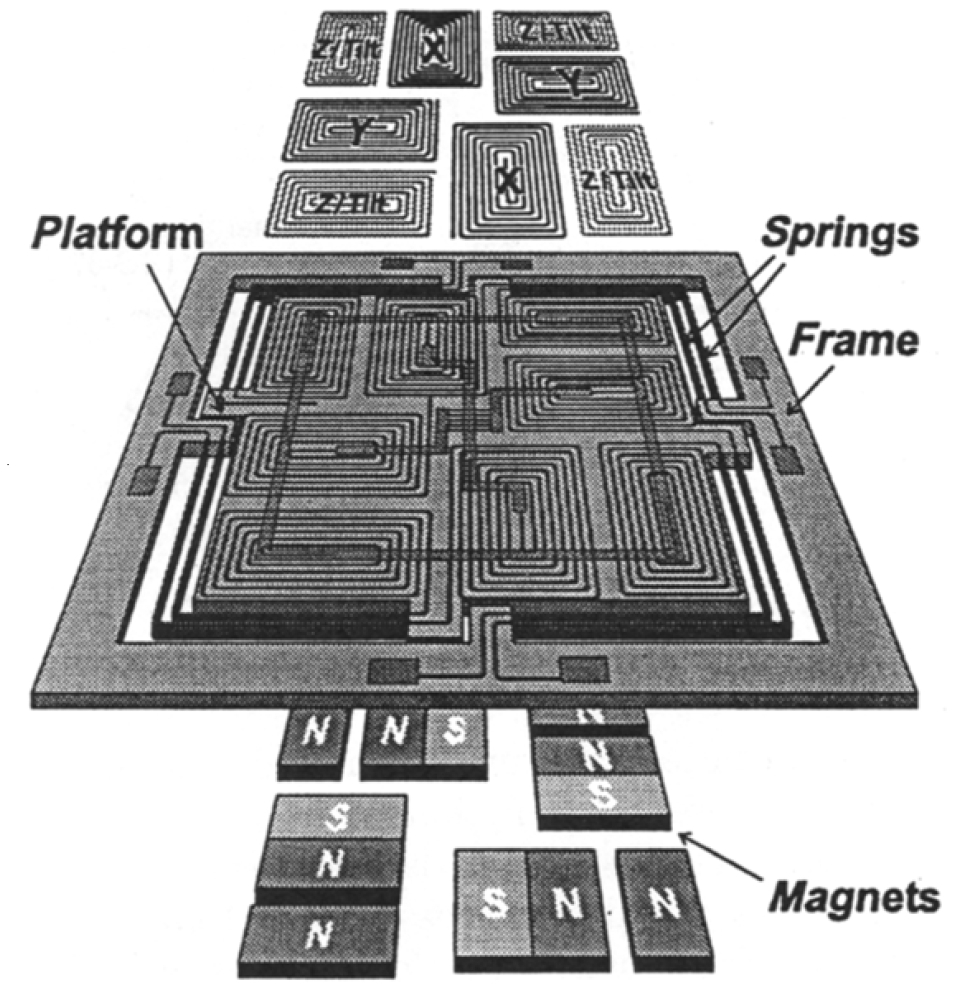}   
  \caption{The initial proof-of-concept electrodynamic scanner by IBM. The scan table is \SI{2x2}{cm} large and has $5$ degrees of freedom. Reproduced with permission from~\onlinecite{Rothuizen2000}. \label{fig:Rothuizen2000}}
\end{figure}

\begin{figure}
  \centering
  \includegraphics[width=7cm]{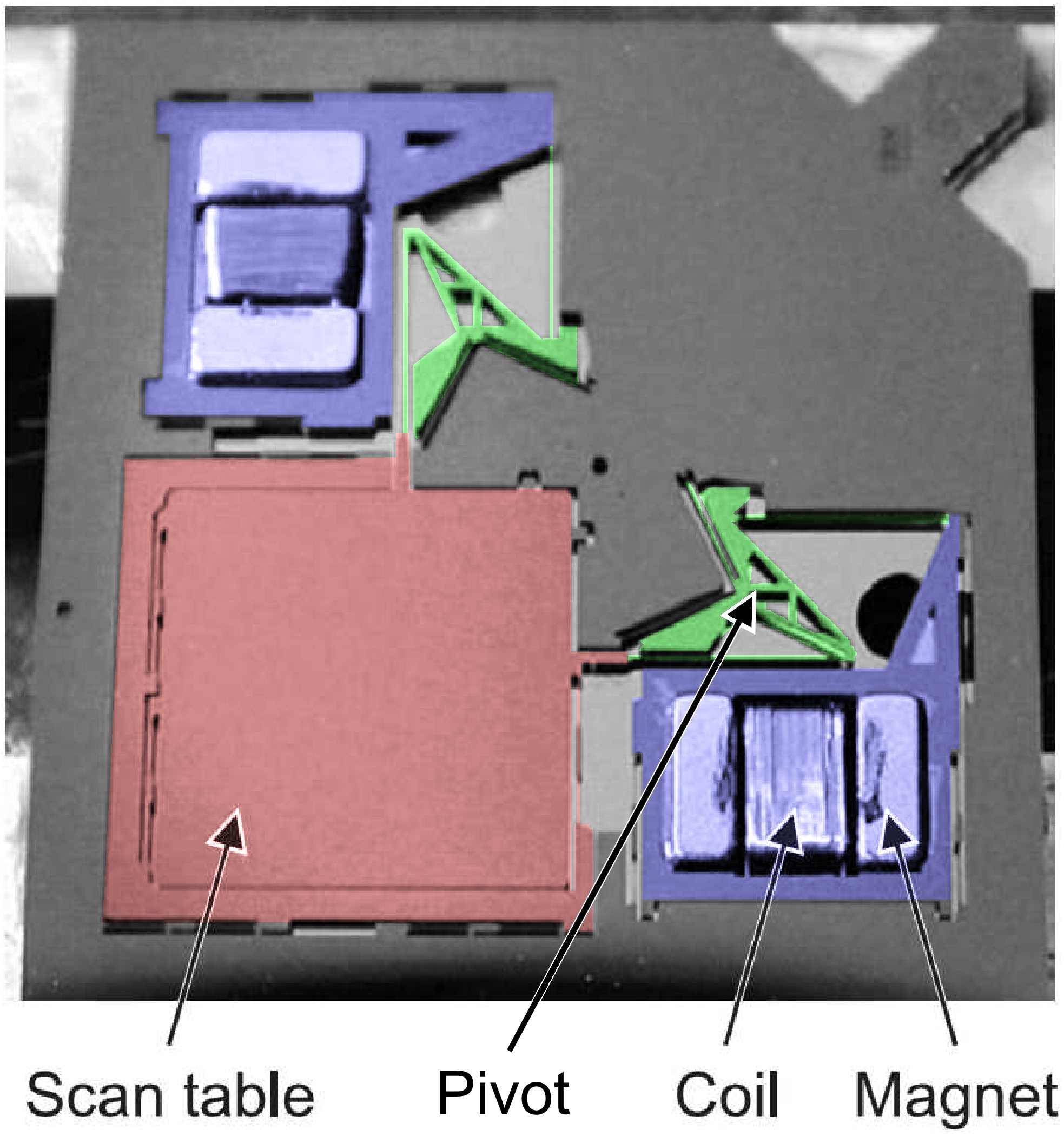}   
  \caption{A mass-balanced electrodynamic 2D scanner by IBM. The scan table size is \SI{6.8x6.8}{mm}. From ref.~\onlinecite{Pantazi2004} \copyright \ IOP Publishing. Reproduced by permission of IOP Publishing. All rights reserved.. \label{fig:Pantazi2004}}
\end{figure}

In 2000, Rothuizen~\etal{} from IBM reported their proof-of-concept electrodynamic scanner for probe data storage (see Figure~\ref{fig:Rothuizen2000}): a five degrees of freedom $x$/$y$/$z$ scanner, including tilt about the $x$- and $y$-axes, fabricated from silicon and electroplated copper springs and coils. 
The scanner contains a \SI{2x2}{cm} moving platform held in a \SI{3x3}{cm} outer frame~\cite{Rothuizen2000}. 
The displacement range is \SI{+-100}{\micm}; however the required power of about \SI{200}{mW} is very high. 
An improved design was reported two years later \cite{Rothuizen2002}, fabricated from a 200-\si{\micm}-thick SU-8 layer, which uses a similar configuration for the coil and magnet. 
It improves on power dissipation (\SI{3}{mW} at \SI{100}{\micm} displacement), fabrication cost, and compactness by placing the spring system below the moving platform. 
Two years later, a radical change in design was reported \cite[see Figure~\ref{fig:Pantazi2004}]{Pantazi2004, Lantz2007}. 
This new design is fabricated from a 400-\si{\micm}-thick silicon wafer by deep reactive-ion etching through the full thickness of the wafer, i.e., the design is an extrusion of a two-dimensional layout. 
It features a mass-balancing concept to render the system stiff against external shocks while keeping it compliant for actuation such that the power dissipation is low.
 The actuator and scan-table masses are linked via a rotation point, enforcing their movement in mutually opposite directions: when the actuator moves up, the table moves down and visa versa.
 External shocks exert inertial forces on the actuator and scan-table mass, but, because the directions of the inertial forces are equal, they cancel each other through the rotation point.
Because the springs are \SI{400}{\micm} high (wafer thickness), the stiffness in the $z$-direction is large for passive shock rejection. 
Coils and magnets are glued manually onto the device. 
The actuator generates a force of \SI{62}{\micro\newton\per\milli\ampere}. 
Its power usage at \SI{50}{\micm} displacement is \SI{60}{mW} (\SI{80}{mA} current); this has been improved to about \SI{2}{mW} (\SI{7}{mA} current) \footnote{Private communication with Mark A. Lantz, IBM Research -- Zurich.}. 
The medium sled is \SI{6.8x6.8}{mm}, while the complete device is \SI{16x16}{mm}; the areal efficiency is about 25\% and has decreased dramatically in comparison to the earlier designs. 
The in-plane resonance frequencies lie around \SI{150}{Hz}; the first out-of-plane resonance frequency lies an order of magnitude higher.

\begin{figure}
  \centering
  \includegraphics[width=9cm]{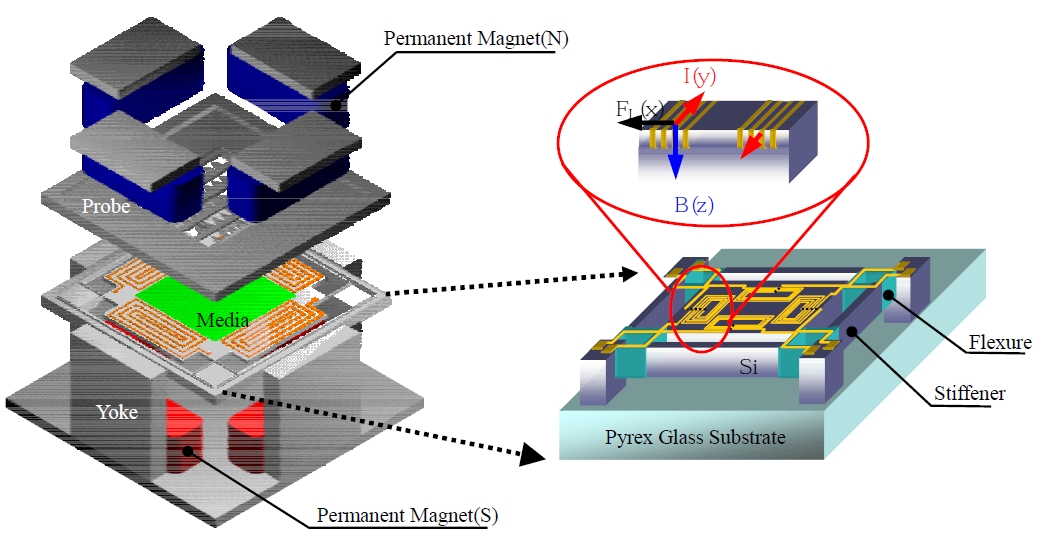}
  \caption{The electrodynamic scanner by Samsung. The size of the total device and scan table is \SI{13x13}{mm} and \SI{5x5}{mm}, respectively. Reproduced with permission from~\onlinecite{Choi2001a}.}
  \label{fig:choi2001a}
\end{figure}

Another electrodynamic scanner was reported in 2001 by Choi~\etal{} from Samsung \cite{Choi2001a, Choi2001b}, see Figure~\ref{fig:choi2001a}. That scanner is fabricated from silicon; the coils are made by filling high-aspect-ratio silicon trenches. The medium sled size is \SI{5x5}{mm}. The displacement of \SI{13}{\micm} at \SI{80}{mA} is smaller than that achieved by the scanner by IBM; however, the displacement was measured without the top magnets and yokes that were planned in the design to increase the magnetic field and force. The measured in-plane resonances are \SI{325}{Hz} (translational) and \SI{610}{Hz} (rotational).

A plastic electrodynamic scanner has been developed by Seagate~\cite{Huang2009}; however, little design and fabrication information is available. The scanner has three degrees of freedom: $x$/$y$ translation and rotation about the $z$-axis. The \SI{+-150}{\micm} displacement range is large, but the resonance frequency is only \SI{70}{Hz}. 

%################################################################################

\subsection{Electrostatic comb-drive actuation}
\label{sct:combdrives}

Electrostatic comb drives \cite{Tang1989} are microelectromechanical actuators that are often used in MEMS technology, because their design and fabrication are relatively straightforward.
A comb drive consists of two interdigitated structures, shaped like combs, that attract each other when they are charged oppositely by application of an external voltage.
The force generated, however, is small, and many fingers are needed to generate sufficient force, reducing the areal efficiency. Other electrostatic actuators, such as the dipole surface drive and the shuffle drive (see Section~\ref{sct:elsteppermotors}), can generate more force, but are more complicated to manufacture and to control.   
The force generated is proportional to and in the direction of increasing capacitance; the combs attract each other because the increase in overlap between the comb fingers increases the capacitance~\cite{Johnson1995a}. 
The capacitance is a function of the spacing (the gap) between fingers, and there is a sideways force on a finger if the finger is not exactly centered between the two opposing fingers. Therefore, the spring suspension must restrict motion in the transverse direction to prevent this `side snap-in'.
A comb drive's force is proportional to the square of the applied voltage.
Because of this quadratic dependence, the polarity of the applied voltage is unimportant, and the force is always attractive. 
To move a stage in both positive and negative direction, two or more comb-drive actuators are required.
Because the spring suspension force depends linearly on the position, the displacement depends quadratically on the applied voltage. 
This relation can be linearised by driving the comb drives in differential mode, reducing the force of one comb drive by simultaneously applying a voltage on the comb drive pulling in the opposite direction \cite{Grade2004}.

The maximum force of a comb-drive actuator depends on the maximum available voltage, if not the breakdown voltage, across the finger gap. The breakdown voltage of comb drives is above \SI{300}{V} for common micrometer-sized gaps \cite{Chen2006}, and is usually higher than the available voltage. To increase the maximum voltage, DC/DC step-up conversion is used. Examples of DC/DC converters specifically designed for MEMS applications are a high-voltage-CMOS design that converts voltages from \SI{3}{V} to \SI{380}{V} using an external coil~\cite{Saheb2007}, and a standard CMOS design that is capable of converting \SI{1.2}{V} to \SI{14.8}{V} without external components~\cite{Hong2003}.

Compared with electromagnetic actuators, an advantage of comb-drive actuators is their ease of fabrication. Moreover, their energy consumption is lower, but for a fair comparison with other scanners, also the energy consumption of the driving (DC/DC conversion) circuitry must be taken into account. 
Although a comb drive is a non-linear actuator, driving one is much simpler than driving the electrostatic stepper motors discussed below.
A comb-drive actuator can simultaneously be used as a position sensor by measuring the capacitance which changes as a function of position~\cite{Chu2003, Pang2009, Huang2010}. The capacitance is measured electrically, and the frequency of the measurement voltage is chosen well above the mechanical resonance frequencies to not interfere with the actuation.
Disadvantages include the generally high voltage required and the low areal efficiency because of the large area needed for comb fingers to generate sufficient force.
The force output highly depends on the minimum gap size that can be fabricated; reduced gap sizes achieved with improved fabrication methods will result in stronger comb-drive actuators.
In practice, the maximum force and voltage are limited by side pull-in because of insufficient spring suspension stiffness~\cite{Zhang2014pullin}.
When the spring suspension is not sufficiently stiff in the in-plane direction perpendicular to the direction of travel, the moving part of the comb drive may snap sideways against the stationary part, generally resulting in terminal failure of the device.
Several suspension designs have been published that improve on the common folded-flexure design~\cite{Tang1989} in order to specifically prevent snap-in while maintaining a large compliance in the desired travel direction~\cite{Olfatnia2013, Krijnen2014}. However, these improved suspension designs substantially increase the footprint of the actuator, reducing their attractiveness for use in a high-density data storage device.
Note that these suspension designs are not restricted to comb drives and can also be used for the other actuator types.

\begin{figure}
  \centering
  \includegraphics[width=7cm]{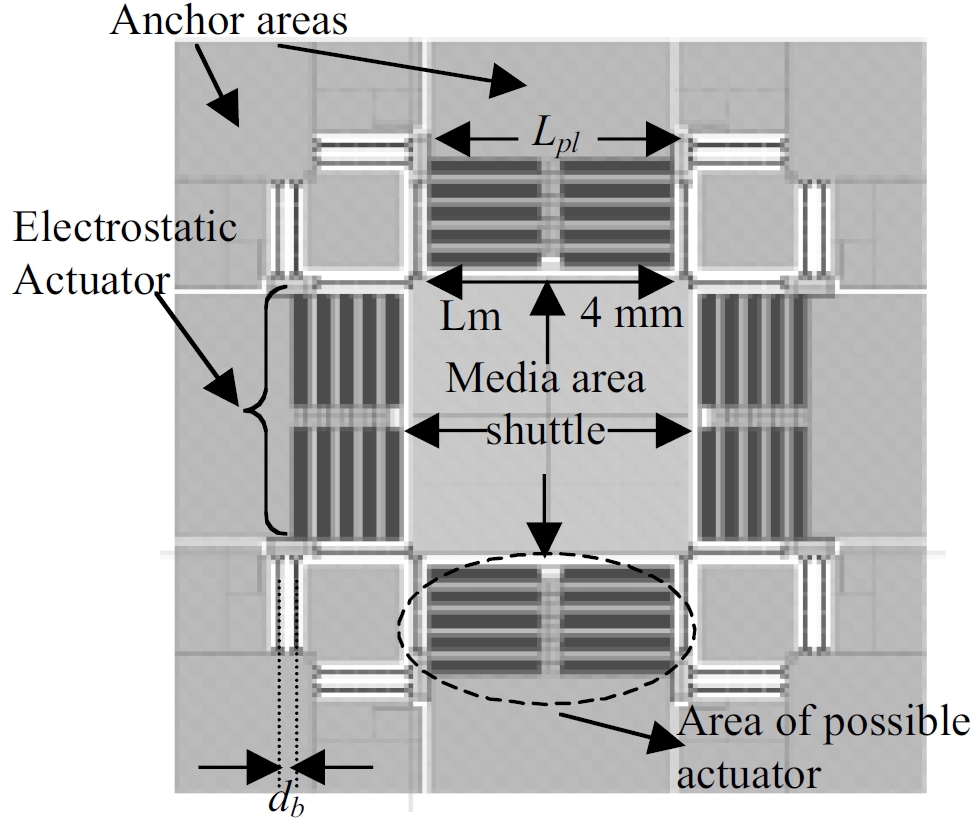}
  \caption{The topology used by Alfaro and Fedder from Carnegie Mellon
    to find a scanner design with optimal footprint for a
    \SI{+-50}{\micm} stroke. Reproduced with permission from~\onlinecite{Alfaro2002}
$^4$.}
%~\protect\footnotemark.}
  \label{fig:Alfaro2002}
\end{figure}

\footnotetext{\copyright 2002 NSTI http://nsti.org. Reprinted and revised, with permission, from the Proc. of International Conference on Modeling and Simulation of Microsystems, pp. 202--205, Apr 22--25, 2002 (San Juan, Puerto Rico, U.S.A.).}

Already in 1992, a nanopositioner with an integrated probe tip was described~\cite{Yao1992}. It featured electrostatic parallel-plate (gap-closing) actuators, moving the probe instead of the `medium'. However, its displacement range of \SI{200}{nm} at \SI{55}{V} is very limited. Another early design included position sensing and control, obtaining a displacement of \SI{8}{\micm} at just \SI{3}{V}~\cite{Cheung1996}. However, the actuator is too thin to carry a load. 

In 2000, Carley~\etal{} from Carnegie Mellon University described a comb-drive scanner for probe storage that had originally been designed for a vi\-bra\-to\-ry-rate gyroscope \cite{Carley2000, Carley2001}.
The scanner reaches \SI{50}{\micm} displacement at \SI{120}{V}; it contains 800 fingers in total, with \SI{500}{\micm} height and \SI{16}{\micm} gap. 
This design was improved using parametric optimization to optimize the footprint for \SI{+-50}{\micm} stroke, keeping the topology fixed~\cite{Alfaro2002} (see Figure~\ref{fig:Alfaro2002}). Although Figure~\ref{fig:Alfaro2002} is not drawn to scale, it is indicative of the generally low areal efficiency of comb-drive scanners.

\begin{figure}
  \centering
  \includegraphics[width=8cm]{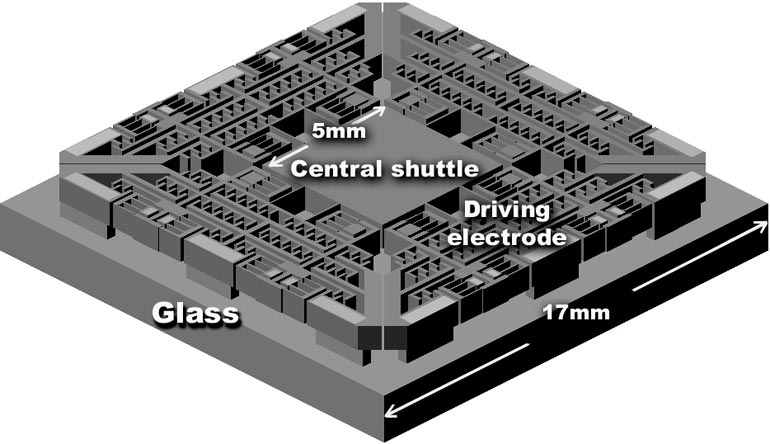}
  \caption{Simplified layout of the micro $x$/$y$-stage by Kim~\etal{} The design features many fingers, generating a large force but reducing the areal efficiency to 11\%. Reproduced with permission from~\onlinecite{Kim2003}.}
  \label{fig:Kim2003}
\end{figure}

The comb-drive scanner reported by Kim~\etal uses 27,552 3-\si{\micm}-thick fingers with \SI{48}{\micm} height and \SI{3}{\micm} gap~\cite{Kim2003} (see Figure~\ref{fig:Kim2003}). These dimensions are common for comb-drive designs, where the maximum aspect ratio (height-to-gap ratio) is limited to approx.~20:1 by the deep reactive-ion etching fabrication process \cite{Jansen2009abrev}.
The scan table is \SI{5x5}{mm} large, and \SI{18}{\micm} static displacement is reached at \SI{13.5}{V}. 
Because of the large number of fingers, the force is large and the required voltage is low; however, this also leads to a low areal efficiency of approximately 11\%.

\begin{figure}
  \centering
  \includegraphics[width=10cm]{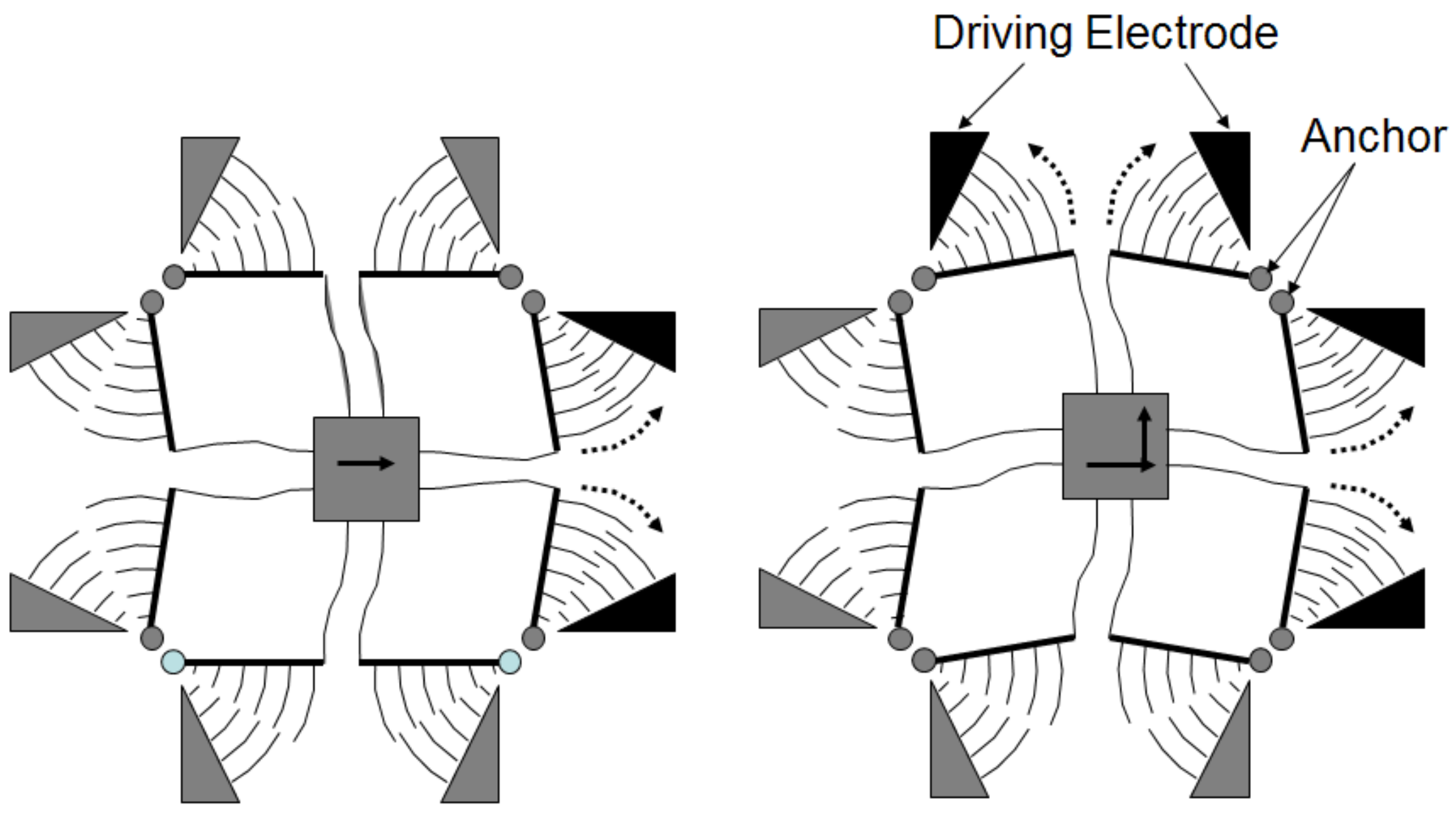}
  \caption{Simplified model of the actuation principle used in the
    design by Kwon~\etal, featuring curved comb
    drives. Left: Movement to the right by actuating the rotational comb
    drives on the right; Right: diagonal movement by actuating the comb
    drives on the right and top. Reproduced with permission from~\onlinecite{Kwon2006}.}
  \label{fig:Kwon2006}
\end{figure}

Kwon~\etal published an $x$/$y$-scanner for optical applications~\cite{Kwon2006}, but the design could also be applied to probe storage. The design uses `L'-shaped suspension springs and rotational comb drives (see Figure~\ref{fig:Kwon2006}). Its displacement range is \SI{55}{\micm} at \SI{40}{V}; however it is unclear whether the suspension is stiff enough for use in probe storage.

\begin{figure}
  \centering
  \includegraphics[width=9cm]{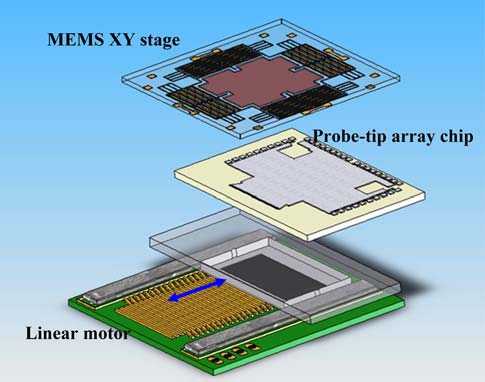}
  \caption{A schematic of the design by the Data Storage Institute in
    Singapore~\cite{Yang2007, Pang2009}, featuring a comb-drive $x$/$y$-stage
    that carries the storage medium, and beneath it a miniature linear
    motor with a 1D probe array. The MEMS stage's size is
    \SI{1x1}{cm} (actuators included). Reproduced with permission from~\onlinecite{Yang2007}.}
  \label{fig:yang2007}
\end{figure}

The somewhat unconventional positioning system design by the Data Storage Institute in Singapore features an electrostatic comb-drive $x$/$y$-scanner \cite{Lu2005} for precise positioning and a 1D miniature electromagnetic linear motor for coarse positioning of a 1D probe array~\cite{Yang2007, Pang2009} (see Figure~\ref{fig:yang2007}).
By using an extra coarse positioner, a small number of probes in a 1D array can be used instead of a 2D probe array.
The comb-drive scanner features a scan table of \SI{6x6}{mm}; the areal efficiency is 36\%. Simulations indicate a static displacement of \SI{20}{\micm} at \SI{55}{V}, but no measurement results are reported. 

All comb-drive scanners mentioned above directly link the actuators to the scan table. This means that in-plane shocks have to be compensated actively for by the actuators. Especially at large displacements, when the available force is low, this leads to a low shock-rejection capability. 
The shape of the comb-drive fingers may be tailored to increase the available force and to provide improved 
shock rejection~\cite{Engelen2009, Engelen2010optim}.
The electrostatic scanner designs by Sasaki~\etal use mass balancing for \emph{internal} shock force rejection~\cite{Sasaki2008}. Inertial forces due to fast acceleration in the $y$-direction may influence the comb finger gap in the $x$-direction, which may lead to instability in the comb drives for the $x$-direction.
To cancel these inertia forces in the $y$-direction, the scan table is split into two plates of equal mass that are actuated in mutually opposite directions. External inertial forces are not cancelled as opposite movement is not mechanically enforced. The best of the three designs investigated reached static displacements of \SI{110}{\micm} at \SI{70}{V} and \SI{90}{\micm} at \SI{125}{V} for the $x$- and $y$-directions, respectively~\cite{Sasaki2008}.

A very direct comparison between comb drives and electrodynamic actuation is the retrofitting of IBM's mass-balanced scanner~\cite{Pantazi2004, Lantz2007} with electrostatic comb drives~\cite{Engelen2008}. The comb-drive scanner uses tapered comb-drive fingers, obtaining displacement ranges of \SI{+-52}{\micm} at \SI{156}{V} and \SI{+-38}{\micm} at \SI{119}{V} in the $x$- and $y$-directions, respectively. The resonance frequencies in both directions (around \SI{140}{Hz}) are slightly lower than those of the electrodynamic scanner.
The total area of the scanner is \SI{3}{mm} larger in both directions than the \SI{16x17}{mm} total area of the electrodynamic scanner. However, because no permanent magnets are needed, the comb-drive version can be made thinner than the electrodynamic scanner.

%################################################################################
\subsection{Electrostatic stepper motors}
\label{sct:elsteppermotors}

\begin{figure}
  \centering
  \subfigure{
    \centering
    \label{fig:Agarwal2006cross}
    \includegraphics[width=4cm]{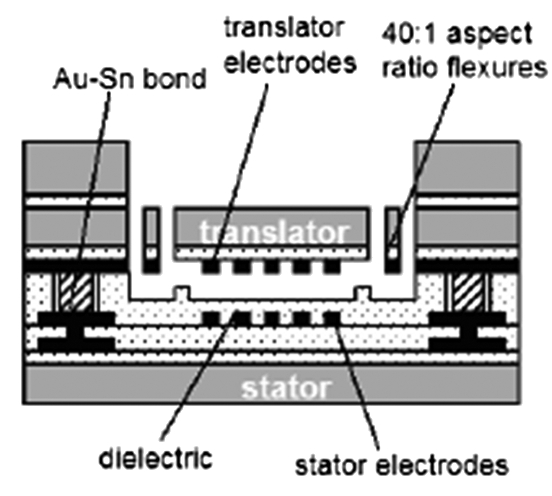}   
  }
  \hspace{1em}
  \subfigure{
    \centering
    \label{fig:Agarwal2006wiring}
    \includegraphics[width=4cm]{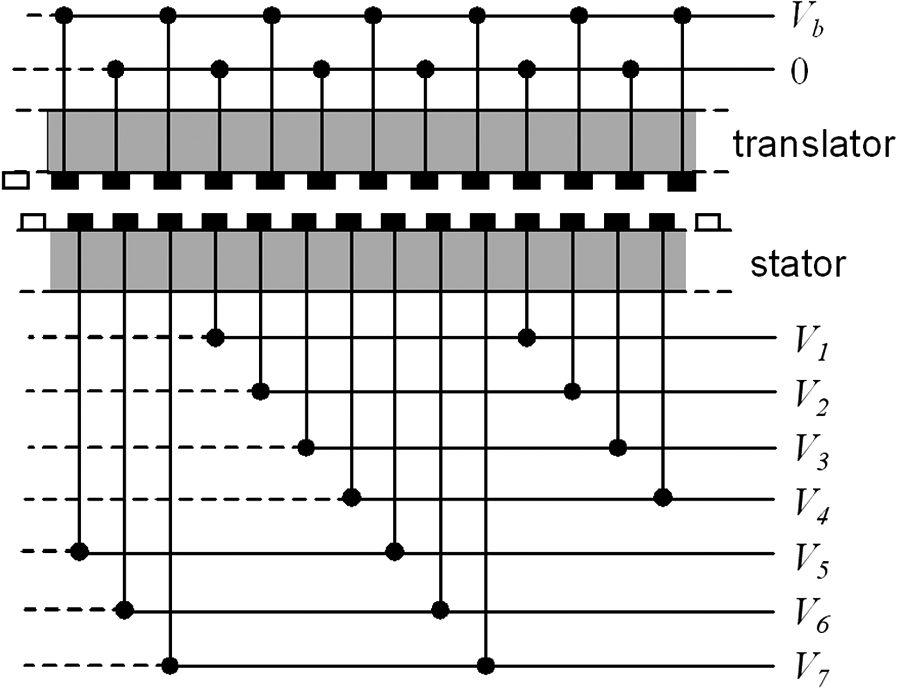}   
  }
  \caption{The dipole surface drive actuator by Agilent. Left, schematic cross section of the dipole surface drive. Right, wiring diagram of the translator and stator electrodes. The translator and stator electrodes have a different pitch, and by sequentially applying a voltage on $V_1$, $V_2$, \ldots, $V_7$, small steps to the left are made by the translator. Reproduced with permission from~\onlinecite{Agarwal2006}.}
  \label{fig:Agarwal2006}
\end{figure}
Figure~\ref{fig:Agarwal2006} shows the electrostatic dipole surface drive of Agilent \cite{Hoen2003,Agarwal2006}. The translator features a periodic pattern of electrodes and is suspended above the stator, which also has a periodic pattern of electrodes but with a different pitch. Oppositely charged translator and stator electrodes will try to align with each other, increasing the capacitance between them, and creating a translational force. 
Sequentially applying a voltage on $V_1$, $V_2$, \ldots, $V_7$, Figure~\ref{fig:Agarwal2006wiring}, results in a stepping motion of the translator to the left.
The step size of \SI{400}{nm} is determined by the difference in electrode pitch between stator and translator. Smaller displacements are made by adjusting the voltage on one of the electrodes.
In the design reported by Agarwal~\etal~\cite{Agarwal2006}, the gap between the stator and translator electrodes is \SI{2.4}{\micm}. A high out-of-plane to in-plane stiffness ratio is required, because the available force is limited by the vertical snap-in voltage (\SI{48}{V} in this case). 
A displacement of \SI{17}{\micm} was reached with \SI{30}{V} bias; larger displacements of up to \SI{70}{\micm} were reached, but in that case significant out-of-plane motion was observed~\cite{Agarwal2006}.
The out-of-plane force can be cancelled by another dipole surface drive placed `upside-down' on top of the translator.

The dipole surface drive can potentially provide a large force at low voltages, because the capacitance (and thus the force) between translator and stator electrodes can be made large owing to the small gap size that is not limited by lithography or deep reactive-ion etching. 
Because the actuator is placed under the scan table, the areal efficiency is greatly increased compared with, e.g., comb-drive designs.
However, the fabrication process and drive circuitry are more complex than for electromagnetic actuators and electrostatic comb drives.
To make the design more shock-resistant, perhaps a 2D mass-balancing scheme could be used similar to IBM's electrodynamic scanner \cite{Lantz2007} without decreasing the areal efficiency.

\begin{figure}
  \centering
  \includegraphics[width=6cm]{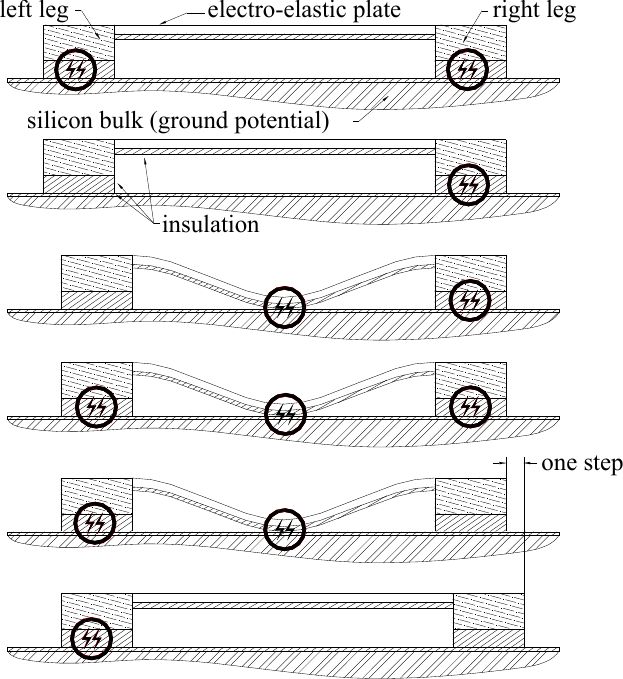} 
  \caption{Step sequence of a shuffle motor. The circles indicate that a voltage is applied on the corresponding leg or plate. Reproduced with permission from~\onlinecite{Patrascu2007b}.}
  \label{fig:Patrascu2007b}
\end{figure}
Another electrostatic stepper motor is the `inchworm' shuffle drive by Tas~\etal~\cite{Tas1998} from the University of Twente. The shuffle drive consists of two clamp `feet' connected by a thin plate. The feet and plate can be pulled towards the bulk individually.
Figure~\ref{fig:Patrascu2007b} shows how the motor makes a step. 
The actuator's output force and step size are determined by how much plate deformation is obtained when applying a voltage on the plate. 
The step size is on the order of tens of nanometers, but highly depends on the displacement because of the restoring force of the spring suspension~\cite{Patrascu2007b}. 
The force generated is large, because the capacitance change of the parallel-plate gap-closing actuator is relatively large and because the plate acts as a mechanical lever.
Sarajlic~\etal describe a \SI{200x1500}{\micm} shuffle motor with a \SI{70}{\micm} displacement range and an output force of \SI{0.45}{mN} at driving voltages of \SI{65}{V} and \SI{150}{V} for the plate and clamps, respectively~\cite{Sarajlic2005d}.
A \SI{482x482}{\micm} 2D shuffle motor reached displacements of \SI{60}{\micm} (corresponding to \SI{0.64}{mN} force) at driving voltages of \SI{45}{V} and \SI{36}{V} for the plate and clamps, respectively, being limited only by the design layout \cite{Sarajlic2005a}. 
Unfortunately, stiction and friction of the feet are a large problem for reliable operation \cite{Patrascu2006}. Moreover, a complex control loop is required \cite{Patrascu2007b}.
Interestingly, because the device is electrostatically clamped to the base plate, 
it is inherently shock-resistant, 
but when the shock force plus the spring force exceed the stiction force, there is no way to compensate for them.

%################################################################################ 
\subsection{Piezoelectric actuation}

Piezoelectric actuation is commonly used for scanning probe microscopy, where there is ample space for the actuator. Piezoelectric materials deform when a voltage is applied, resulting in a translational, rotational, or bending motion.
Piezoelectric elements need to be quite large to provide the required displacement range for probe storage~\cite{Muralt2000}. 
Fortunately because the generated force is large, mechanical stroke amplification can be used to increase the displacement range of small piezoelectric elements. 
Commonly required voltages are comparable to the voltages required for electrostatic actuators. 

\begin{figure}
  \centering
  \includegraphics[width=8cm]{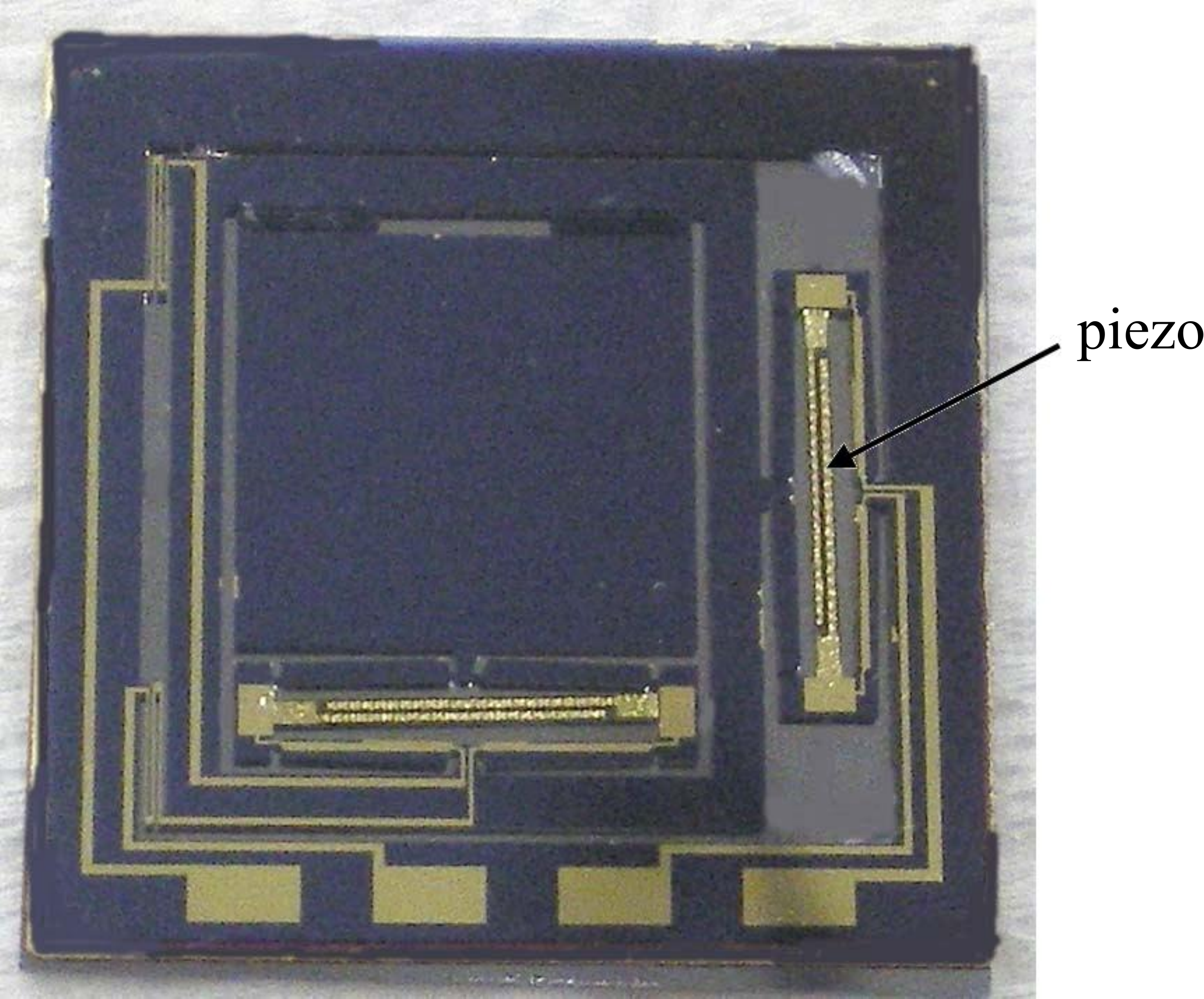}
  \caption{Photograph of the silicon microstage by Faizul~\etal to which stacked PZT actuators have been attached. The total device is 2$\times$2~cm$^2$ in size. Reproduced with permission from~\onlinecite{Faizul2009}.}
  \label{fig:faizul2009}
\end{figure}
Several scanner designs using PZT actuators and stroke amplification are described by Zhang~\etal~\cite{Zhang2003}. An improved design reached \SI{82}{\micm} at an applied voltage of \SI{70}{V} \cite{Faizul2009b}. That scanner consists of a silicon MEMS displacement stage to which PZT actuators have been attached manually (see Figure~\ref{fig:faizul2009}). The scan table is \SI{9.5x9.5}{mm} in size. 
Although the fabrication of a micromachined stacked PZT actuator especially designed for the scanner is discussed, only measurements using larger commercial PZT actuators are shown~\cite{Faizul2009b}.
The PZT actuators generate large forces ($\sim$\SI{3}{N}) and their small displacement is mechanically amplified 20 times to move the scan table. The maximum force of \SI{150}{mN} on the scan table is more than an order of magnitude larger than those of the scanners discussed above.
The displacement versus voltage measurements show hysteresis, but this can be reduced by using charge control instead of voltage control \cite{Comstock1981}.
It will be interesting to see whether the stiff suspension provides a passive shock-resistance comparable to the shock resistance of IBM's mass-balanced electrodynamic scanner. 
The fabrication complexity of attaching the PZT actuators on the silicon stage is comparable to the glueing of permanent magnets onto IBM's electrodynamic scanner shown in Figure~\ref{fig:Pantazi2004}.

%################################################################################

%#################################################################
\section{Probe arrays and parallel readout}
\label{sec:probe-array}

%\subsection{Introduction}
Although the probes used in probe-based data storage are originally derived from standard AFM and STM probes, they have become very complex over time. Not only do the probes require electrical actuation and readout, they also have to be extremely wear resistant and have to fit into a restricted area. Another challenging task is to realize probe arrays with thousands of tips in parallel. Readout of these arrays is far from trivial, especially as the number of probes increases.

\subsection{Probe technology and arrays}
The most advanced probe arrays have been realized by the IBM--Research
Zurich probe-storage team. Already in 1999, Lutwyche~\etal{}
realized a 5$\times$5 array of probes with tip heaters and
piezoresistive deflection readout~\cite{Lutwyche1999}. Ultrasharp tips
were obtained by oxidation sharpening of isotropically etched
tips. The tips were located at the end of cantilevers that are bent
towards the medium by purposely introducing stress gradients to clear
the lever anchors from the medium. Boron implantation in specific
regions of Si cantilevers was used to define piezoresistors and tip
heaters. To increase the sensitivity up to a $\Delta R$/$R$ of \SI{4e-5}{nm^{-1}}, where $R$ is the resistance, constrictions were introduced at the base of the cantilever, but these constrictions led to higher resistance, increasing the pink noise.

In the subsequent 32$\times$32 array, piezoresistive sensing was abandoned~\cite{Despont2000}. In this array, the cell size was reduced from \SI{1000}{\micm} to \SI{92}{\micm}, while keeping the cantilever spring constant at \SI{1}{N.m^{-1}} with a resonance frequency of \SI{200}{kHz}. The array size was 3 mm $\times$ 3 mm, and thermal expansion deteriorating the tip alignment became an issue. Integrated heaters were positioned in the array to keep temperature variations within \SI{1}{\celsius} over the chip. The array worked remarkably well: \SI{80}{\%} of the cantilevers worked~\cite{Lutwyche2000} and a density of \SI{200}{Gb.in^{-2}} at \SI{1}{Mb.s^{-1}} net data rate was shown~\cite{Durig2000}.

\begin{figure}
  \centering
  \includegraphics[width=\figurewidth]{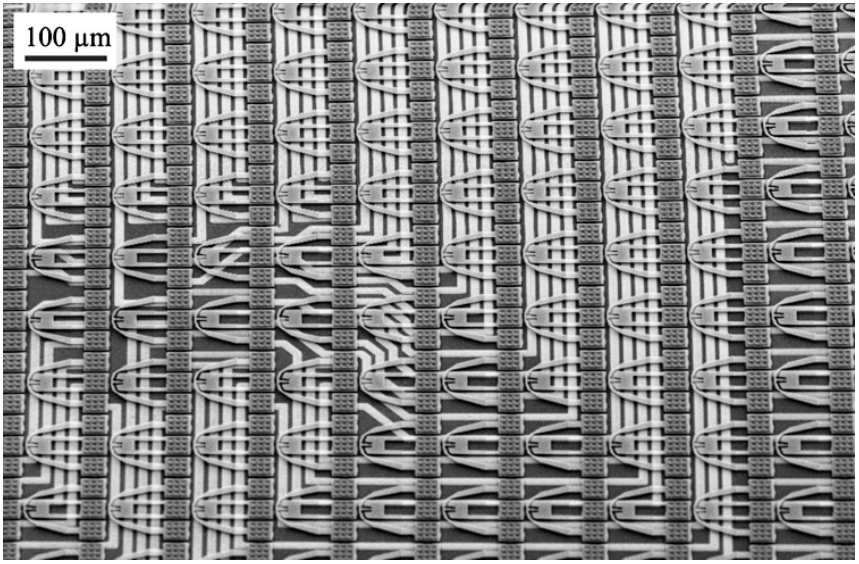}
  \caption{SEM image of part of a 4096 cantilever array, interconnected to a wiring wafer. Reproduced with permission from~\onlinecite{Despont2004}.}
  \label{fig:Despont2004}
\end{figure}

An impressively tight integration of the probe array with CMOS was demonstrated by Despont~\etal~\cite{Despont2004}. In this method only the integrated cantilevers are transferred to the CMOS chip, and the MEMS carrier wafer is first ground and then etched away. On \SI{1}{mm^2}, as many as 300 electrical copper interconnects of \SI{5}{\micm} were realized. An array of 4096 probes with outer dimensions of 6.4 mm $\times$ 6.4 mm was constructed (Figure~\ref{fig:Despont2004}), and the interconnects had a yield of \SI{99.9}{\%}.

As tip wear can be reduced by applying less force to the tip, the probe design was modified so that the spring constant reduced to \SI{0.05}{N.m^{-1}}. As a result, the probe applies very little force to the tip during read actions. During write actions, this force can be electrostatically increased up to \SI{1}{\micro\newton} by means of a capacitive platform at a potential of \SI{20}{V}.

% LG
The work of IBM triggered the interest of other companies. For heated tip writing on piezoelectric and phase-change media, researchers at LG Electronics in Korea built a small array of thermal probes~\cite{Lee2002a}. Heater platforms were integrated into boron-doped silicon by creating a constriction at the cantilever end and covering the cantilever legs with gold. Conductive tungsten tips were grown by focused ion-beam deposition. In the next generation, readout was added by integrating piezoelectric PZT layers on the cantilever legs~\cite{Lee2003}. Feature heights of \SI{30}{nm} could easily be distinguished.
The array was extended to a size of 128$\times$128 probes, and the sensitivity improved to \SI{20}{nm}~\cite{Nam2007}. A wafer-transfer method was developed for a 34$\times$34 array~\cite{Kim2006,Kim2007}, very much along the lines of the IBM process. 
Rather than silicon, $300$-nm-thick silicon nitride probes were used with integrated polysilicon heaters. The spring constant was still relatively high (\SI{1}{N.m^{-1}}). Sharp tips were realized by KOH etching of pits into the silicon wafer and subsequent filling of the pits with silicon nitride, enabling bit dimensions of \SI{65}{nm}.

%Twente
At the University of Twente, cantilever arrays with small variations in tip-sample distance were fabricated~\cite{Koelmans2011a}. As tip wear increases with load force, it is important that the tip-sample distance is similar for each tip in an array. The well-known KOH etching process for single tips, reported in~\cite{Albrecht1990}, was extended to arrays of cantilevers that feature self-aligned tips. The simple and easily scalable fabrication method was demonstrated in an array of 10 tips that showed a standard error of \SI{10}{nm} in tip-sample spacing~\cite{Koelmans2011a}.

% Let op, bao 2 papers of array
Researchers at the Shanghai Institute of Microsystems and Information Technology have realized a small cantilever array, with integrated heater tips and piezoelectric deflection detection~\cite{Yang2006}. These arrays have been used to characterize the wear of polymer recording media as a function of tip temperature and radius~\cite{Bao2008}.
 
% Data storage insitute
At the Data Storage Institute in Singapore, Chong~\etal{} realized 20$\times$1 and 15$\times$2 arrays, using a fabrication technique along the lines of the early IBM work~\cite{Chong2005}. The scanning concept, however, is 
different, and features a large stroke actuator in one direction. The total storage area can thus be much larger than the size of the array~\cite{Yang2007}. The $90$-\SI{}{\micm}-long and $1$-\SI{}{\micm}-thick cantilevers of the 
array had a spring constant of \SI{1}{N~m^{-1}} and a resonance frequency of \SI{164}{kHz}. The tip radius was rather large (\SI{220}{nm}), resulting in bit dimensions on the order of \SI{600}{nm}. 
Cooling times of \SI{2}{\micro\second} were measured by monitoring the heater platform temperature with an infrared camera. 
%\johan{reference needed}

\begin{figure}[!hbt]
  \centering
  \includegraphics[width=7cm]{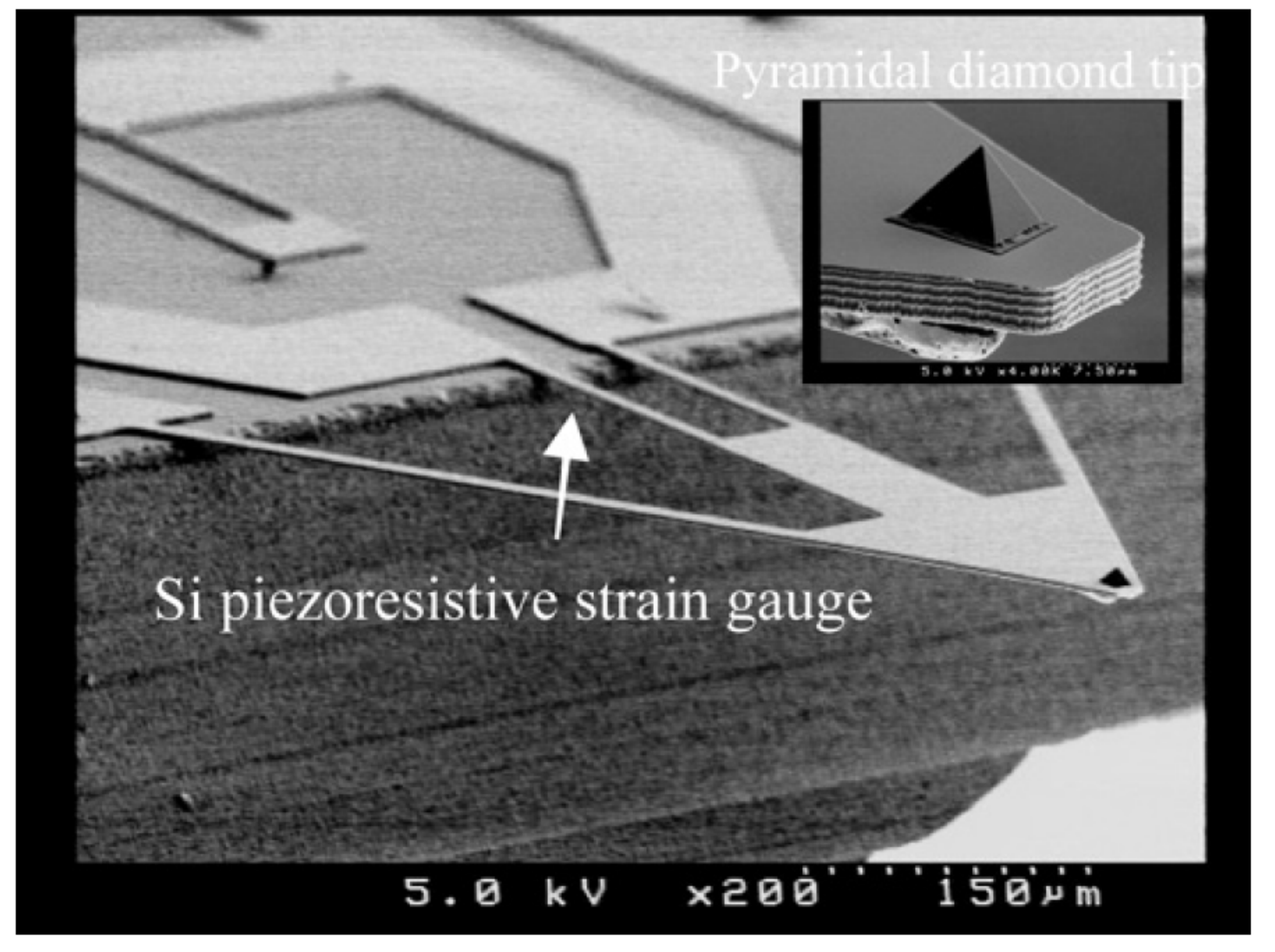}
  \caption{Diamond probe with a silicon-based piezoresistive strain gauge. From ref.~\onlinecite{Takahashi2006} \copyright \ IOP Publishing. Reproduced by permission of IOP Publishing. All rights reserved.}
  \label{fig:Takahashi2006}
\end{figure}

% Pioneer, Ferroelectric, Takahashi
%\cite{Takahashi2007,Takahashi2004,Takahashi2006,Takahashi2006a}
Researchers at Pioneer and at the Tohoku University in Japan
investigated arrays of probes with diamond tips and integrated
piezoresistive sensors for ferroelectric data
storage~\cite{Takahashi2006a}, see Fig.~\ref{fig:Takahashi2006}. The
boron-implanted Si piezoresistive Wheatstone bridge was very sensitive,
with a $\Delta R$/$R$ of \SI{1e-7}{nm^{-1}}~\cite{Takahashi2006}. In contrast, the diamond probes had a relatively poor radius of curvature. Attempts were made to replace the diamond tips by metal versions, and it was demonstrated that Ruthenium tips perform relatively well~\cite{Takahashi2007}.

\subsection{Parallel operation}
\label{sec:par-readout}

The massive parallelism of the probe arrays, which is prerequisite for obtaining data rates comparable to those of magnetic hard-disk heads, poses a large challenge in probe-based data storage. Several thousands of probes have to be addressed simultaneously. 
The main functions of each probe are positioning, reading and writing. Positioning, as described in section \ref{sct:positioning}, is in most cases done by moving the complete array or the storage medium in plane and parallel to the probe array, simplifying the task of each probe.

\label{scanningthemedium} Scanning the medium has two distinct advantages over scanning the probes. To obtain desired read and write speeds, arrays have to scan at considerable rate, thereby inducing high-frequency vibrations that create unwanted cantilever movement. 
Preventing the occurrence of these vibrations is a major additional challenge for any control loop. Secondly, electrical connections to the probes can be realized more easily, because the probes do not move with respect to the read-channel electronics. 
Researchers at the Data Storage Institute presented a solution where the coarse-positioning stage has a flexible wire to the readout electronics, although also in this design the fine positioning is directly connected to the cantilever array~\cite{Yang2007}.

Movement in the $z$-direction, where $z$ is defined as the direction
normal to the medium, can be done on a per-array basis instead of on a
per-probe basis. This hugely simplifies the control required to
operate an array of thousands of probes. However, the fabrication
tolerances of the array and the medium have to be such that every probe in
the array is in the appropriate tip-medium distance range. Too large a
tip-sample separation results in a failure when an attempt to write or
read a bit is done. The other extreme leads to a probe that is pushed
into the medium with considerable force (depending on the spring
constant of the cantilever), leading to excessive tip wear. Without
independent $z$-motion, these demands on the medium and probe array
increase tremendously. The technically most mature probe-storage
system, described in~\cite{Pantazi2008}, is based on thermomechanical
storage and features a 64$\times$64 cantilever array, of which $32$ are
active. By determining the electrostatic pull-in voltage for each
cantilever, the initial tip-sample separation is calculated to have a
standard deviation of \SI{180}{nm}. With a cantilever spring constant
of \SI{1}{N.m^{-1}}, this would lead to an additional loading
force of hundreds of nN.

The read and write operations require the independent addressing of each probe. Traditionally, AFM probes are monitored by an optical readout system, mainly based on the optical beam-deflection technique~\cite{Meyer1988, Meyer1988er}. Although optical readout has been demonstrated for arrays of cantilevers~\cite{Lang1998,Sulchek2001,Alvarez2005, Koelmans2010, Koelmans2011thesis, Pjetri2011,Honschoten2011} none of the readout schemes has been implemented in probe storage. Optical readout alone, however, would not suffice. The probes have to be actuated for the write operation. Wired schemes are implemented to achieve this. Wireless schemes and passing signals through the storage medium have been proposed~\cite{Abelmann2003}, but not yet realized.

\begin{figure}
  \centering
  \includegraphics[width=9cm]{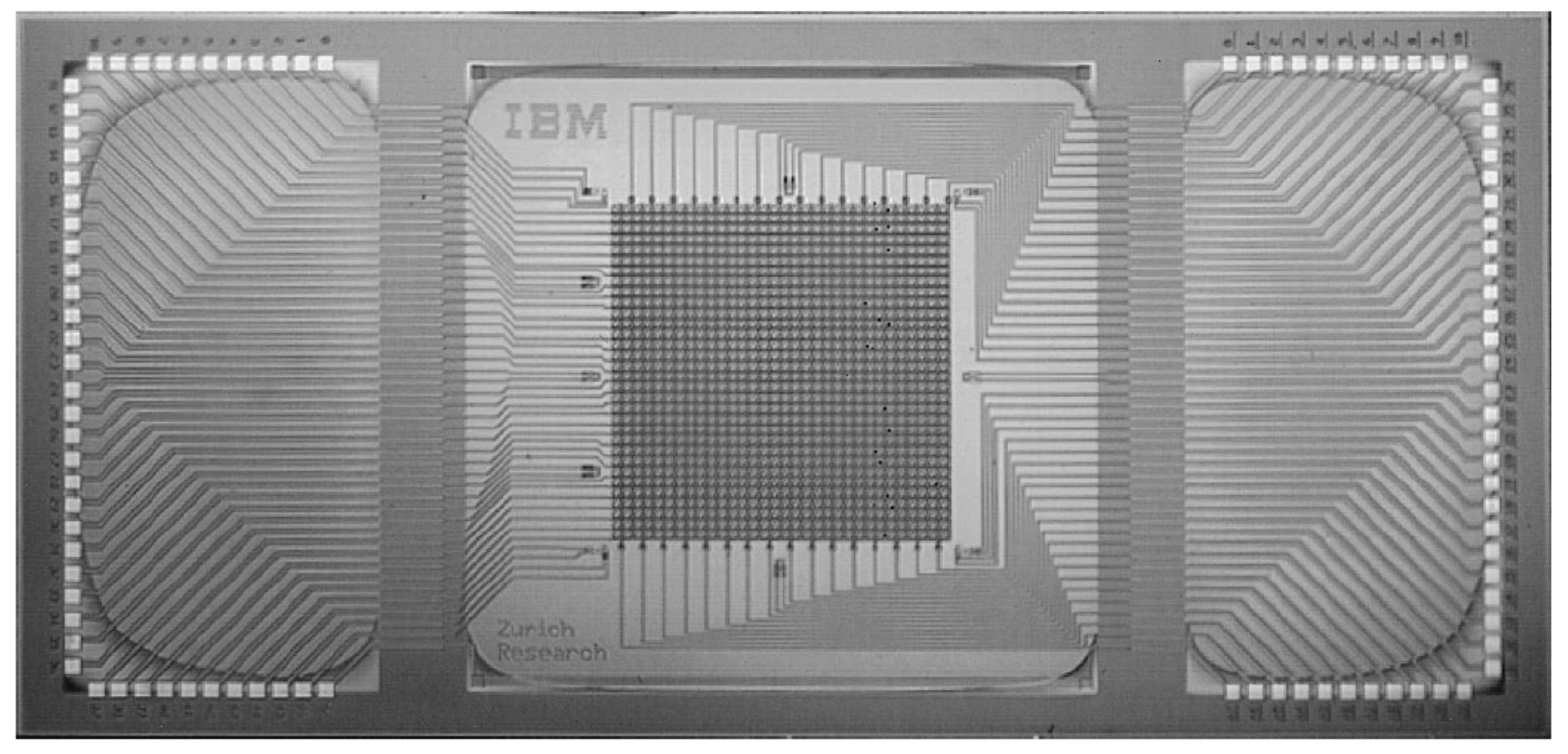}
  \caption{Photograph of a 32$\times$32 array (14$\times$7~mm$^2$). The 
  probes are interconnected by a 32$\times$32 row/column addressing 
  scheme. 128 bond pads are visible on the left- and right-hand side of the array. Reproduced with permission from~\onlinecite{Despont2000}.}
  \label{fig:Despont2000}
\end{figure}

Wiring solutions are based on a time-multiplexing scheme to address the array row by row~\cite{Despont2000,Kim2007}, as is usually done in DRAM. With a growing number of probes, the maximum current passed through a row or signal line increases. For higher numbers of probes, an electrically stable wiring material, capable of carrying high current densities and also having a low resistivity, must be used. 
As shown in Figure~\ref{fig:Despont2000}, Despont and co-workers have used two-level wiring of both gold and nickel in the 32$\times$32 array~\cite{Despont2000}. Schottky diodes formed by doped silicon/nickel interfaces were introduced to avoid crosstalk between probes.

\begin{figure}
  \centering
  \includegraphics[width=9cm]{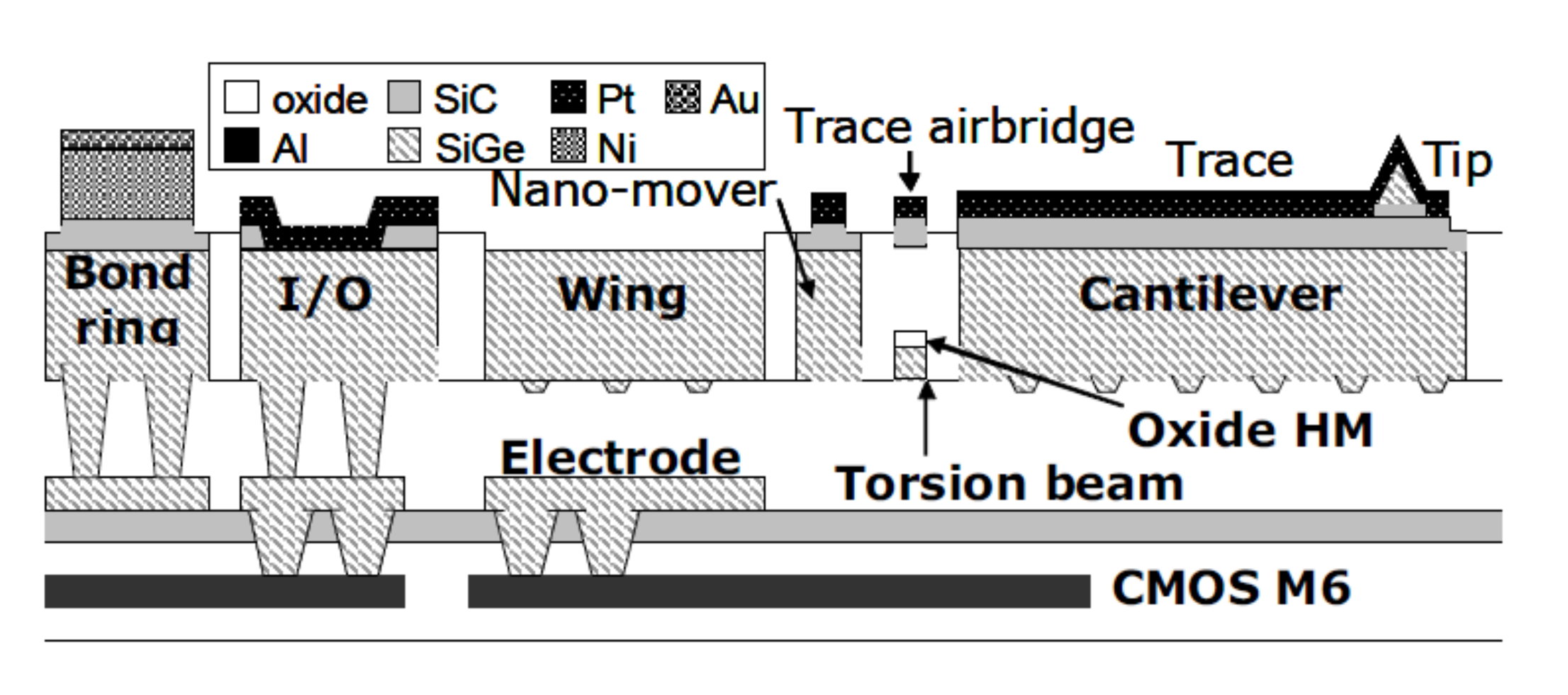}
  \caption{Schematic drawing of a cross section showing the integration of a 
  cantilever array on top of CMOS circuitry, based on a one-wafer design. Reproduced with permission from~\onlinecite{Severi2009}.}
  \label{fig:Severi2009}
\end{figure}

In the 64$\times$64~array reported in~\cite{Despont2004}, the
number of connections has increased to three per probe. In this case,
a second wafer with the CMOS circuitry is used to which the probes are
bonded, as done in~\cite{Kim2007}. An alternative integration
with CMOS is the single-wafer process described
in~\cite{Severi2009}. Here first the CMOS circuitry is created, 
and on the last metalization layer, a chemical mechanical
polishing (CMP) step is performed.
Next an insulating layer of $400$-nm-thick silicon carbide and a sacrificial
oxide layer of \SI{3}{\micm} are deposited. A structural silicon
germanium layer of \SI{3}{\micm} is later deposited to allow
definition of the cantilevers. Then the cantilevers are etched and the
sacrificial oxide is removed to allow cantilever actuation, see Figure~\ref{fig:Severi2009}.
%%% Local Variables:
%%% mode: latex
%%% TeX-master: "PrStReview"
%%% End:

%#################################################################
\section{Final remarks}
Probe-based data storage has been a very active area of research,
leading to a prototype demonstration in 2005 using topographic storage. Topographic probe-storage is the most mature probe-storage technology. Very high densities of \SI{4}{Tb in^{-2}}, have been achieved in combination with excellent retention, endurance, and read and write speeds. 

Phase-change probe storage is severely limited by the difficulty 
of amorphizing crystalline regions of the storage medium. Most of 
the work on phase-change probe storage has been discontinued, and the 
results feed into the currently ongoing realization of solid-state 
phase-change memory cells. Probes have successfully demonstrated 
the scalability of phase-change memory and allow rapid nanoscale 
characterization.

Magnetic probe storage was jump-started by hard-disk research, 
leading to many demonstrations of and insights into the 
switching of bit-patterned media, which are currently an active topic of research in the hard-disk industry. Demonstrations are limited to a density of a few \SI{10}{Gb/in^2}. Such densities give rise to the question whether magnetic probe storage should be pursued any further. 

Ferroelectric storage shows promising densities, read speeds and endurance numbers, 
and has attracted the interest of academia and industry. 
%However, the amount of research that was published is not so large and industry has abandoned pursueing ferroelectric probe storage.

The research into probe storage has led to huge contributions to probe
and array technologies, medium development, positioning, and sensory
and control systems for the nanometer scale. In probe and array
technology, much work was done scale up the single probe used in
scanning probe microscopy to an array of probes reading and writing
data. Wear of probe and media have been greatly
reduced.
The most prominent positioning systems are MEMS scanners using electrodynamic (using a coil and a permanent magnet) or electrostatic comb-drive actuators, achieving sub-nanometer positioning accuracy.
The mature positioning system of the 2005 prototype demonstration uses electrodynamic actuators with a mechanical design for shock-resistance, bringing it closer to productization.
Probe-storage research has strongly impacted and contributed to a
variety of other application areas, e.g., probe-based nanolithography,
probe-based nanofabrication, 3D probe patterning and fast nano-scale
imaging.

No commercial probe-storage product exists as of today. One of the reasons that major efforts by industry to develop a probe-storage system have stopped is the
accelerated introduction of Flash memory, driven by the introduction
of the digital cameras and the USB memory stick. 

Mechanically addressed storage does offer a cost advantage over lithographically-defined storage. For mechanically-addressed storage, the relatively large cost of the initial system dominates the total device cost, whereas the media cost is relatively low. In contrast, electrically-addressed storage has little initial overhead cost, and the total cost is dominated by the capacity. 
Therefore, scaling to more media per device (adding platters in an HDD) or using exchangable media (tape storage) is significantly more cost-effective than increasing the die area of a Flash memory chip.
This is why the magnetic
hard-disk remains the nonvolatile storage system of choice for
data centers in terms of capacity, and why tape storage is a thriving industry for data storage at even larger capacities.
For a probe-based storage device, we envision a cost structure similar to that of the other mechanically-addressed storage devices.
Moreover, mechanically-addressed storage systems are much less dependent on the lithographic capabilities to scale the data density, e.g., chemical etching techniques are used to reach the probe dimension required and the probe, in turn, creates very high data densities on a relatively cheap medium.

Current commercial storage technologies have reached such scale that the cost has been reduced to a point where other performance characteristics besides the areal density, such as low access time, high readback data rate, and small form factor, strongly influence the choice of technology.
However, to reach atomic densities or atomic nano-manipulation capability, the probe-based approach is clearly the best suited and most mature technology.

The authors hope this review will provide future
researchers a quick and comprehensive insight into the amazing
achievements in probe-based data storage that paved the way for 
future nanotechnology research and applications.

%%% Local Variables:
%%% mode: latex
%%% TeX-master: "PrStReview"
%%% End:

%#################################################################
\section*{Acknowledgments}
We gratefully acknowledge our (former) colleagues in the $\mu$SPAM team at the University of Twente for the many stimulating discussions that laid the basis for this work. We thank M.~Gemelli, M.~Khatib and O.~Zaboronski for discussion and valuable input. We also thank C.~Bolliger and M.~Lantz for proofreading the manuscript, A.~Sebastian, H.~Bhaskaran and D.~Wright for fruitful discussions and E.~Eleftheriou for his support of this work.

%\section*{References}
%\bibliographystyle{apsrev4-1}
%\bibliographystyle{aipnum4-1}
\bibliography{C:/Data/SVN/PaperBase/paperbase}

\end{document}